\documentclass[review,12pt]{elsarticle}

\usepackage{setspace}
\usepackage{subcaption}
\usepackage{multirow}
\usepackage{multicol}
\usepackage[table,xcdraw]{xcolor}
\usepackage{graphicx}
\RequirePackage{amsmath}
\usepackage{bm}
\usepackage{xfrac}
\usepackage{float}
\usepackage{arydshln} 

\usepackage{amsmath,esint,amsfonts}

\usepackage[most]{tcolorbox}
\usepackage[colorinlistoftodos]{todonotes}

\usepackage{lineno}
\usepackage[colorlinks=true,linkcolor=blue,urlcolor=blue,citecolor=blue,anchorcolor=blue]{hyperref}
\usepackage{cleveref} 

\journal{Journal}

\biboptions{sort&compress} 

\begin{document}
\begin{frontmatter}

	\title{Surrogate Neural Network Model for Sensitivity Analysis and Uncertainty Quantification of the Mechanical Behavior in the Optical Lens-Barrel Assembly}

	\author{Shantanu Shahane\fnref{Corresponding Author}$^{a,b,}$}
	\author{Erman Guleryuz$^{a}$}
	\author{Diab W Abueidda$^{a}$}
	\author{Allen Lee$^{d}$}
	\author{Joe Liu$^{d}$}
	\author{Xin Yu$^{d}$}
	\author{Raymond Chiu$^{d}$}
	\author{Seid Koric$^{a,b}$}
	\author{Narayana R Aluru$^{a,b,c}$}
	\author{Placid M Ferreira$^{b}$}
	\address{a National Center for Supercomputing Applications,\\
		University of Illinois at Urbana-Champaign, Urbana, IL 61801, USA}
	\address{b Department of Mechanical Science and Engineering,\\
		University of Illinois at Urbana-Champaign, Urbana, IL 61801, USA}
	\address{c Walker Department of Mechanical Engineering,\\ The University of Texas at Austin, Austin, TX, 78712, USA}
	\address{d Foxconn Interconnect Technology Limited,\\ 2001 S Fifth Street, Suite 201, Champaign, IL, 61820, USA}
	\fntext[Corresponding Author]{\vspace{0.3cm}Corresponding Author Email Address: \url{shahaneshantanu@gmail.com}}



	%

	\begin{abstract}
		Surrogate neural network-based models have been lately trained and used in a variety of science and engineering applications where the number of evaluations of a target function is limited by execution time. In cell phone camera systems, various errors, such as interferences at the lens-barrel and lens-lens interfaces and axial, radial, and tilt misalignments, accumulate and alter profile of the lenses in a stochastic manner which ultimately changes optical focusing properties. Nonlinear finite element analysis of the stochastic mechanical behavior of lenses due to the interference fits is used on high-performance computing (HPC) to generate sufficient training and testing data for subsequent deep learning. Once properly trained and validated, the surrogate neural network model enabled accurate and almost instant evaluations of millions of function evaluations providing the final lens profiles. This computational model, enhanced by artificial intelligence, enabled us to efficiently perform Monte-Carlo analysis for sensitivity and uncertainty quantification of the final lens profile to various interferences. It can be further coupled with an optical analysis to perform ray tracing and analyze the focal properties of the lens module. Moreover, it can provide a valuable tool for optimizing tolerance design and intelligent components matching for many similar press-fit assembly processes.
	\end{abstract}


\end{frontmatter}


\section{Introduction}

Manufacturing tolerance design and analysis which involves determining the size and location (relative to nominal) of acceptable uncertainty zones in the dimensions of functional features, is an essential and critical design step in improving product quality, reducing overall costs, and retaining market share \cite{chase1991survey}. These tolerance zones are designed to obtain an acceptable balance between functional performance and manufacturing costs of the product. In assembled products such as camera lens modules for smart phones, which are produced in high volumes with sets of miniature (millimeter size) assembled components held together by interference fits, tolerance design for mating features becomes increasingly critical. This is because (i) the process capabilities of typical high-volume manufacturing processes result in uncertainty zones that are large fractions of small nominal dimensions, and (ii) as a result, the strain-fields generated by the dimensional mismatch of interference mating features can influence the geometry of all the functional surfaces of the components in the assembly. While process uncertainties can be controlled with higher precision tooling and more stringent process controls, this control comes at a cost that has an exponential relationship with precision levels \cite{taniguchi1983current}.

There have been several approaches to tolerance design, using optimization techniques that balance manufacturing costs with product performance measures. For example, \citet{lewis1994robust} uses second-moment theory to determine tolerance zones that produce the desired levels of functional reliability while Turner and Wozny \cite{turner1990relative,turner1987tolerances} uses linear programming techniques to arrive at tolerance zones. \citet{chase1991survey} provide a comprehensive survey of the different approaches to tolerance design. The taxonomy provided in the paper is relevant today, because much of the work that has followed are been variations on the approaches identified.

Tolerance design remains a difficult problem, not because of a dearth of models, but because these decisions must be made before actual production begins or before the availability of actual production information. All the methods cited above, and currently used, must make assumptions about costs and performance functions, and their interaction through the uncertainties in critical dimensions, typically using simplifying assumptions about them. A critical and consequential decision such as tolerance design would be better served through a more realistic and physics-guided relationship between tolerance zones and performance. Therefore, this paper seeks to quantify the effects of uncertainty on mating features of components on performance (here error in the lens profiles) through the use of machine-learning models and physics-based computational experimentation.

Machine learning techniques have lately achieved important accomplishments in wide areas of science and engineering, such as in natural language processing, voice recognition, computer vision, medical diagnostic and autonomous vehicle driving. In physics-based numerical modeling, design and optimizations, various surrogate deep learning data-driven models have been devised and trained to learn and quickly inference the thermal conductivity \cite{rong2019predicting}, inverse design of advanced composite manufacturing \cite{goli2020chemnet}, near optimal topologies of meta-materials and structures \cite{abueidda2020topology, kollmann2020deep}, fatigue of materials \cite{spear2018data}, nonlinear material response such as in plasticity and viscoplasticity \cite{mozaffar2019deep, abueidda2021deep}, quantum computing for mechanics \cite{mielke2019evaluating}, and many other similar computational challenging applications. Several successful studies have been also reported in using physics informed deep learning models \cite{raissi2019physics, guo2020solving, abueidda2021meshless} to directly solve the partial differential equations (PDEs), governing some of these the physical laws and processes. Neural networks have also been used as surrogate models coupled with computational methods for sensitivity analysis \cite{shahane2021sensitivity}, uncertainty quantification \cite{shahane2019uncertainty, shahane2019finite, feng2019machine, sankaran2015impact, olivier2021bayesian}, inverse problems \cite{puzyrev2019deep} and design optimization \cite{shahane2020optimization, shahane2019numerical}.

\section{Process Framework}
\begin{figure}[H]
	\centering
	\includegraphics[width=\textwidth]{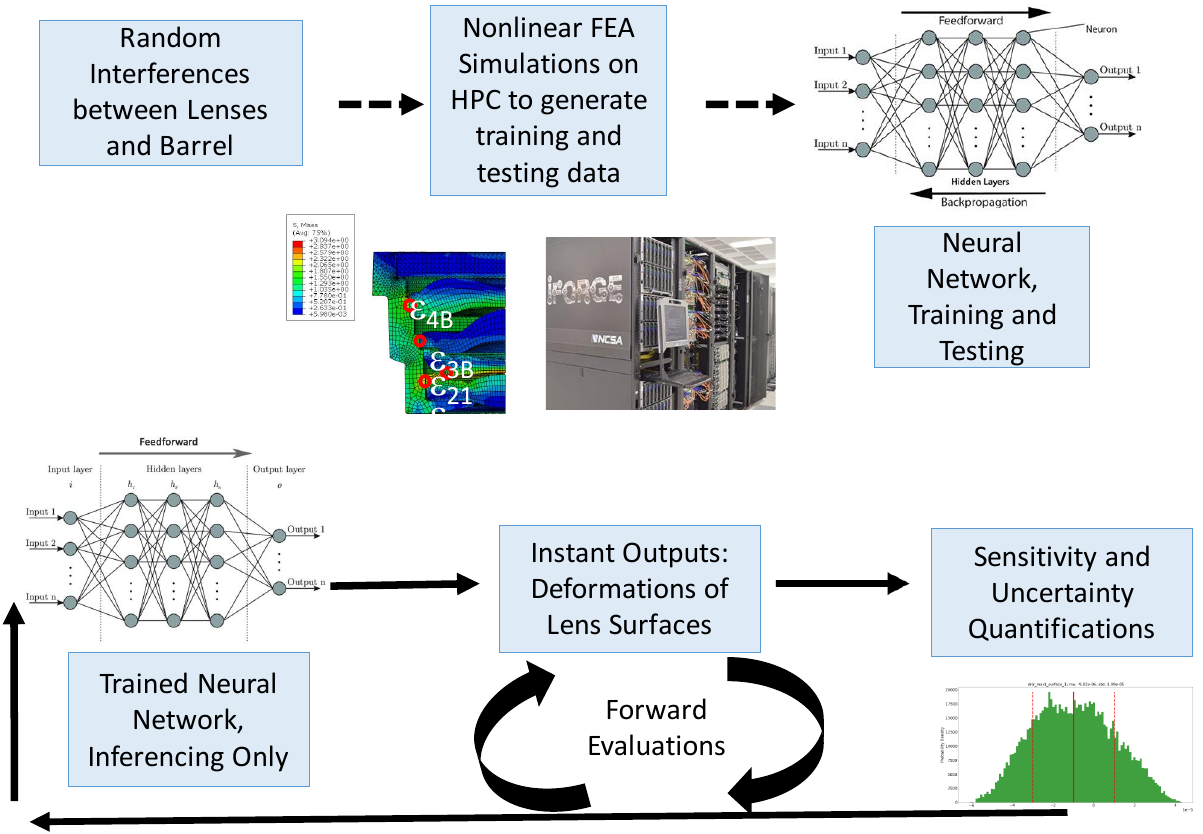}
	\caption{Schematic}
	\label{Fig:overall_schematic}
\end{figure}
\Cref{Fig:overall_schematic} shows the schematic of the overall process. The randomly sampled interferences between lenses and the barrels are inputs to the nonlinear high-fidelity finite element simulation, which provides the deformation and stress in the lenses. Since sensitivity analysis and uncertainty quantification are prohibitively expensive due to the vast number of forward-model numerical evaluations needed to obtain converging statistics, we have first generated training and testing data on HPC. We have then adequately trained and tested a neural network to estimate the deformations from the given interferences. Finally, as a surrogate model, the trained neural network is able to instantly provide millions of accurate forward evaluations required for sensitivity and uncertainty quantification analysis.

\section{Surrogate Model of Lens Assembly for Sensitivity and Uncertainty Analyses}

\subsection{Numerical Model of Lens Assembly Deformation}
\begin{figure}[H]
	\centering
	\begin{subfigure}[t]{0.45\textwidth}
		\includegraphics[width=\textwidth]{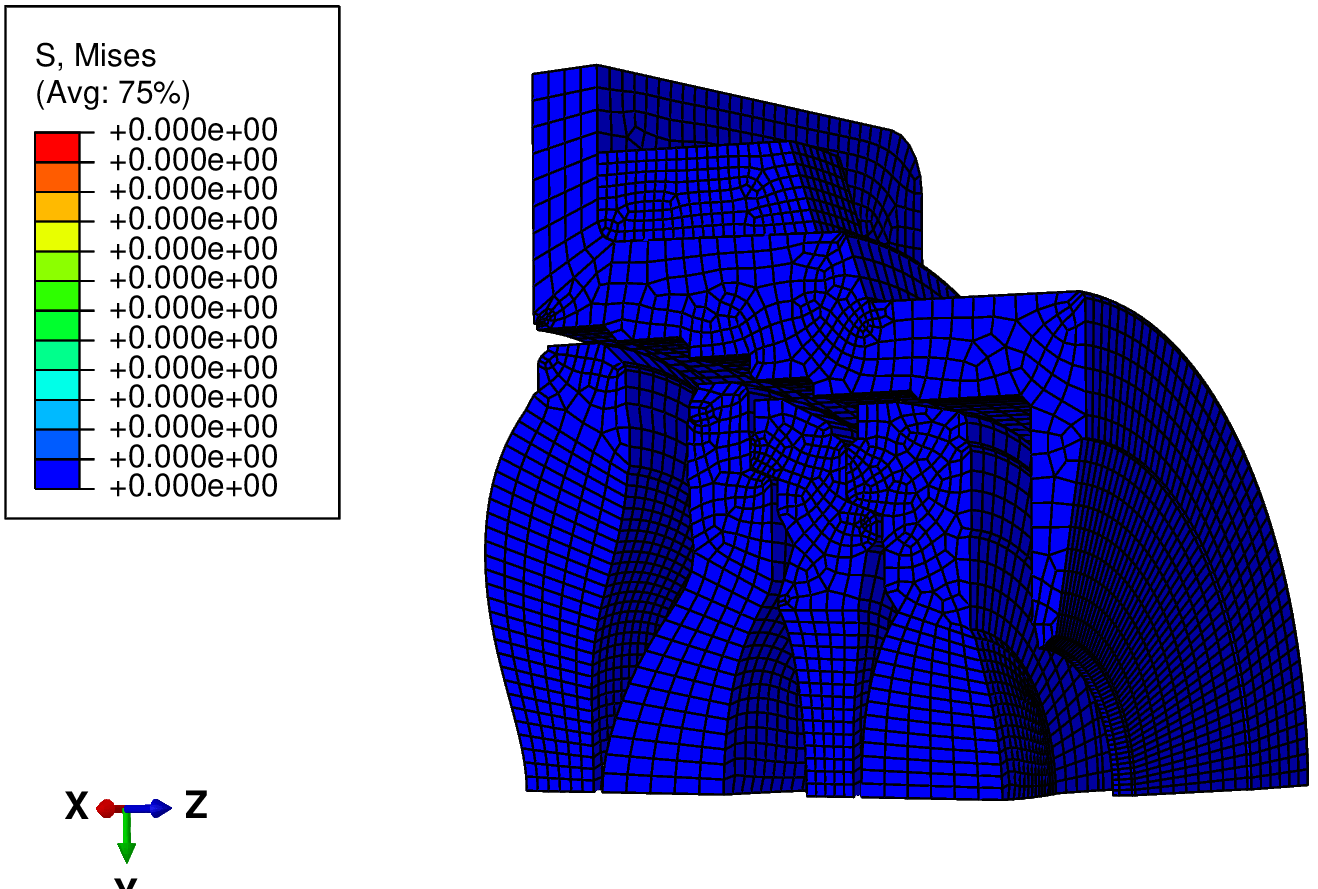}
		\caption{Undeformed Assembly}
	\end{subfigure}
	\hspace{0.05\textwidth}
	\begin{subfigure}[t]{0.45\textwidth}
		\includegraphics[width=\textwidth]{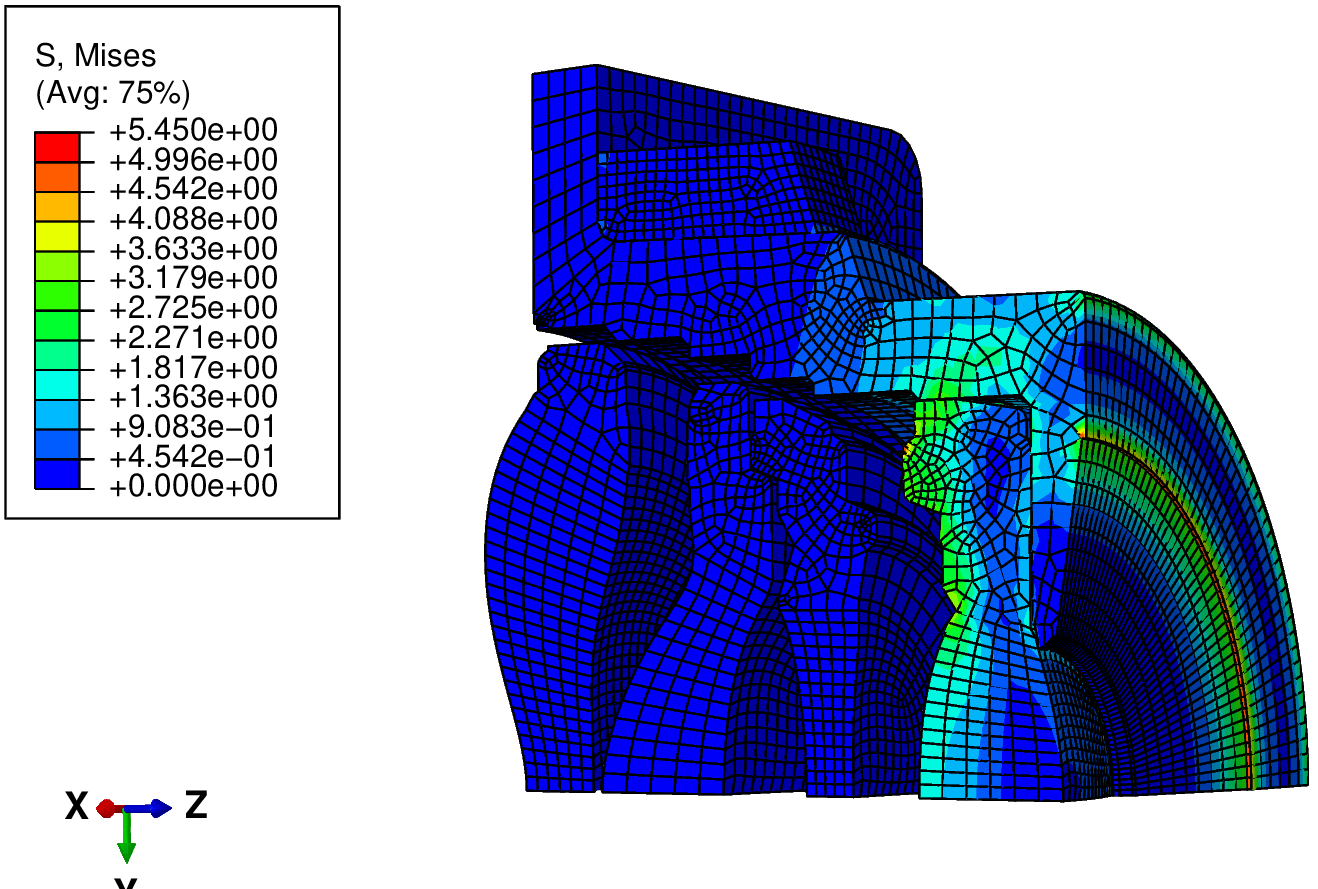}
		\caption{First Lens Assembled} \vspace{0.25cm}
	\end{subfigure}
	\begin{subfigure}[t]{0.45\textwidth}
		\includegraphics[width=\textwidth]{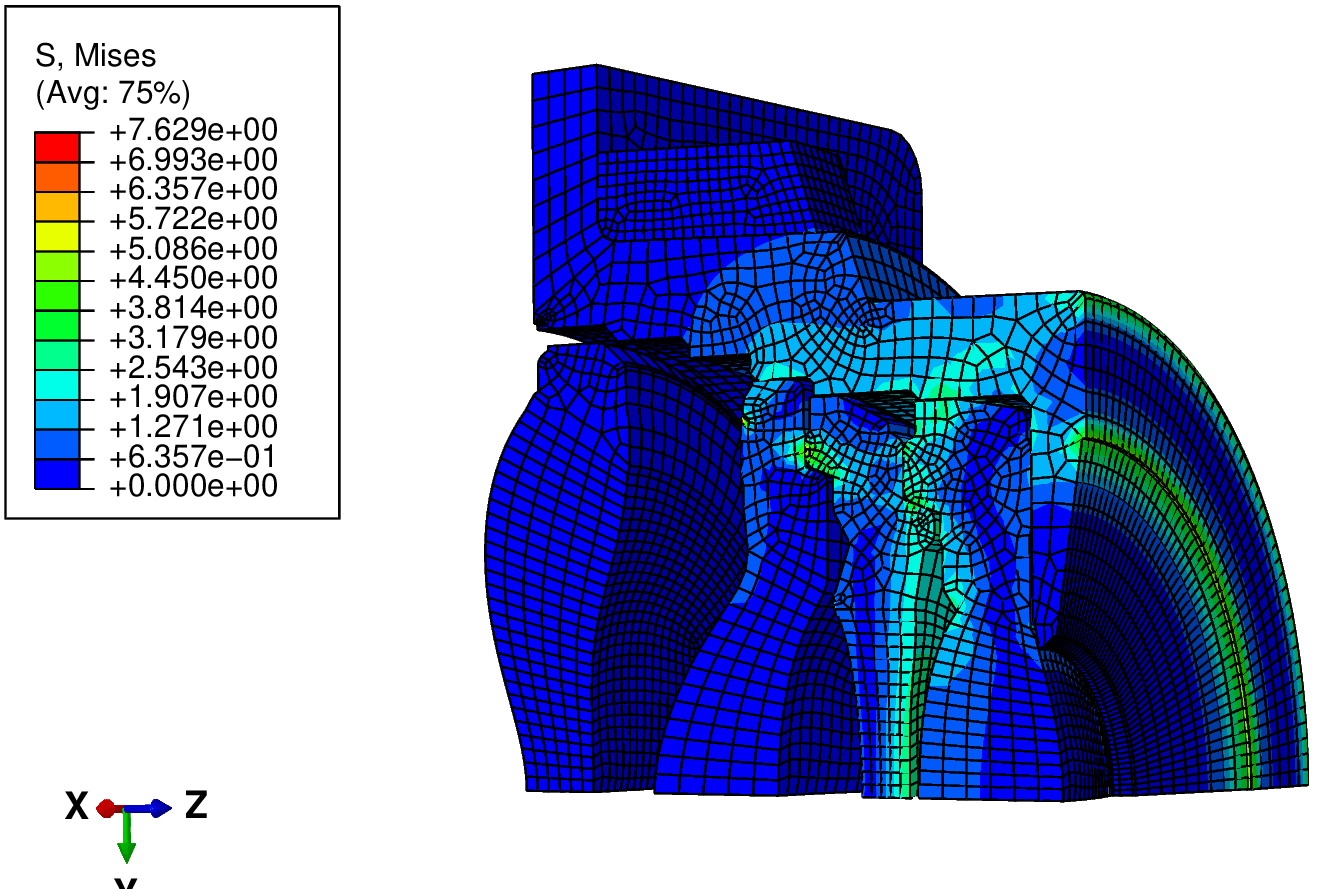}
		\caption{More Lenses Assembled}
	\end{subfigure}
	\hspace{0.05\textwidth}
	\begin{subfigure}[t]{0.45\textwidth}
		\includegraphics[width=\textwidth]{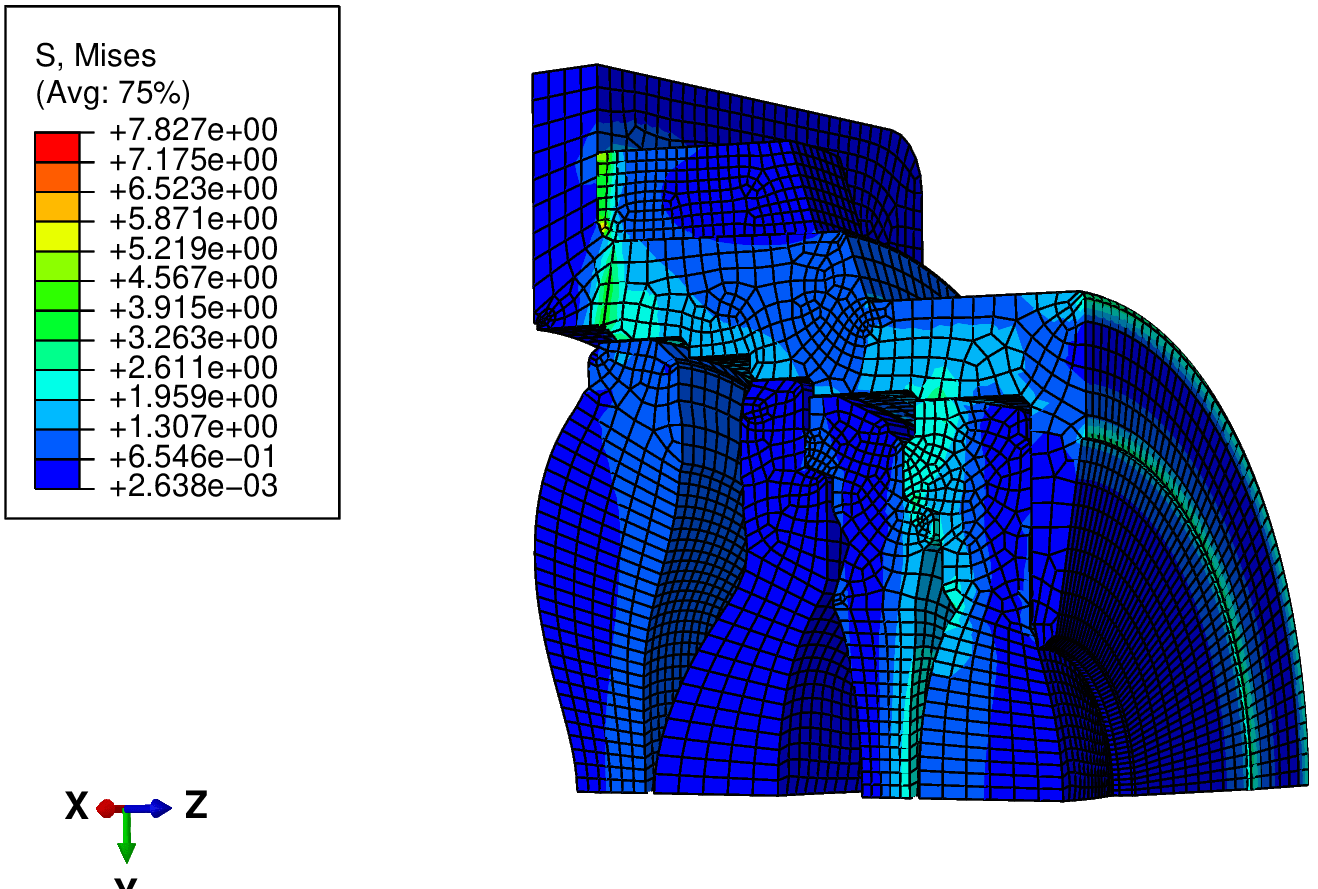}
		\caption{All Lenses Assembled}
	\end{subfigure}
	\caption{Evolution of Deformation Results and Interferences in Multistep Analysis of Lens Assembly}
	\label{Fig:abaqus stress}
\end{figure}

An optical lens module, consisting of four lenses, is assembled in a barrel with precise axial and radial positions to function optically successfully. The components in the lens assembly are manufactured to some specified tolerances that inherently possess a degree of variation and uncertainty in the dimensions, geometry/shape, and relative position of their mating features. A high fidelity implicit finite element model of a quarter of the assembly with symmetry conditions is built to analyze the effect of the interference fits between the components during the assembly. A multistep analysis in \cref{Fig:abaqus stress} is performed, displacing one lens at a time in the optical z-direction to its appropriate place in the assembly. At the same time, the entire barrel structure is kept constrained. The analysis enabled precisely capturing progressing deformation and interferences between the components due to evolving contact interactions during assembly.

The computational model exhibits geometric nonlinearity due to load-induced deformations as well as boundary nonlinearity due changes in contact conditions during analysis. In the implicit finite element formulation used in this work \cite{abaqus_manual}, the nonlinear equilibrium equations of a structure can be represented in their general form, at quasi-static time $t+\Delta t$, as:
\begin{equation}
\bm{p}^{t+\Delta t} - \bm{f}^{t+\Delta t}=0
\label{Eq:abaqus eq 1}
\end{equation}
where $\bm{p}$ (the vector of external loads) and $\bm{f}$ (the vector of internal forces) must balance each other. An iterative approach is needed for the solution of \cref{Eq:abaqus eq 1}, as the internal forces (created by stresses in the elements) depend nonlinearly on the displacements. The Newton-Raphson approach is utilized to find the equilibrium solution using an incremental-iterative procedure. In this approach, the solution is obtained by a sequence of quasi-static time increments ($\Delta t$), with iterations to obtain equilibrium within each increment. The numerical algorithm can be stated as \cite{ibrahimbegovic2009nonlinear}:
\begin{equation}
\bm{r}_i = \bm{p}_i^{t+\Delta t} - \bm{f}_i^{t+\Delta t}=0
\label{Eq:abaqus eq 2}
\end{equation}

\begin{equation}
\bm{K}_i^{t+\Delta t} \Delta \bm{u}_{i+1}  = \bm{r}_i
\label{Eq:abaqus eq 3}
\end{equation}

\begin{equation}
\bm{u}_{i+1}^{t+\Delta t} = \bm{u}_i^{t+\Delta t} + \Delta \bm{u}_{i+1}
\label{Eq:abaqus eq 4}
\end{equation}

\begin{equation}
\bm{u}_{0}^{t+\Delta t} = \bm{u}^{t}
\label{Eq:abaqus eq 5}
\end{equation}

where, $\bm{r}$ is out-of-balance residual force vector, $\bm{u}$ is the vector of displacements at element nodes, $\bm{K}$ is the tangential stiffness matrix of the structure upon an infinitesimal increase of loading, and subscript $i$ indicates the iteration number.
\par The residual vector in \cref{Eq:abaqus eq 2} is first calculated according to the initial configuration. The linear system in \cref{Eq:abaqus eq 3} can then be solved for the displacement increment vector ($\Delta \bm{u}$), usually by a direct sparse solver. For each time increment, displacement increments are accumulated in \cref{Eq:abaqus eq 4} through equilibrium iterations until the convergence is achieved. In nonlinear problems the force residual will never be exactly zero, so it compared to a tolerance value and the iteration is terminated when the Euclidean norm of the residual vector converges to a small tolerance value. The initial conditions are based on a converged solution at the previous time (or load) increment as given in \cref{Eq:abaqus eq 5}. The complete quasi-static loading path is traced with suitably chosen consecutive time increments that lead to convergence. In addition, the contact algorithm in an outer loop determines current contact state at each contact point at each increment, and accordingly imposes constrains based on the Lagrange multiplier method \cite{hallquist1985sliding}, updates K, and performs an equilibrium iteration in \cref{Eq:abaqus eq 2,Eq:abaqus eq 3,Eq:abaqus eq 4}. If the assumed contact state during the equilibrium iteration changes from open to close or vice versa, a new outer loop, so called severe discontinuity iteration (SDI) initiates, repeating the entire process until there is no change in contact conditions and with the mechanical equilibrium satisfied within the convergence criteria.
\begin{figure}[H]
	\centering
	\includegraphics[width=0.6\textwidth]{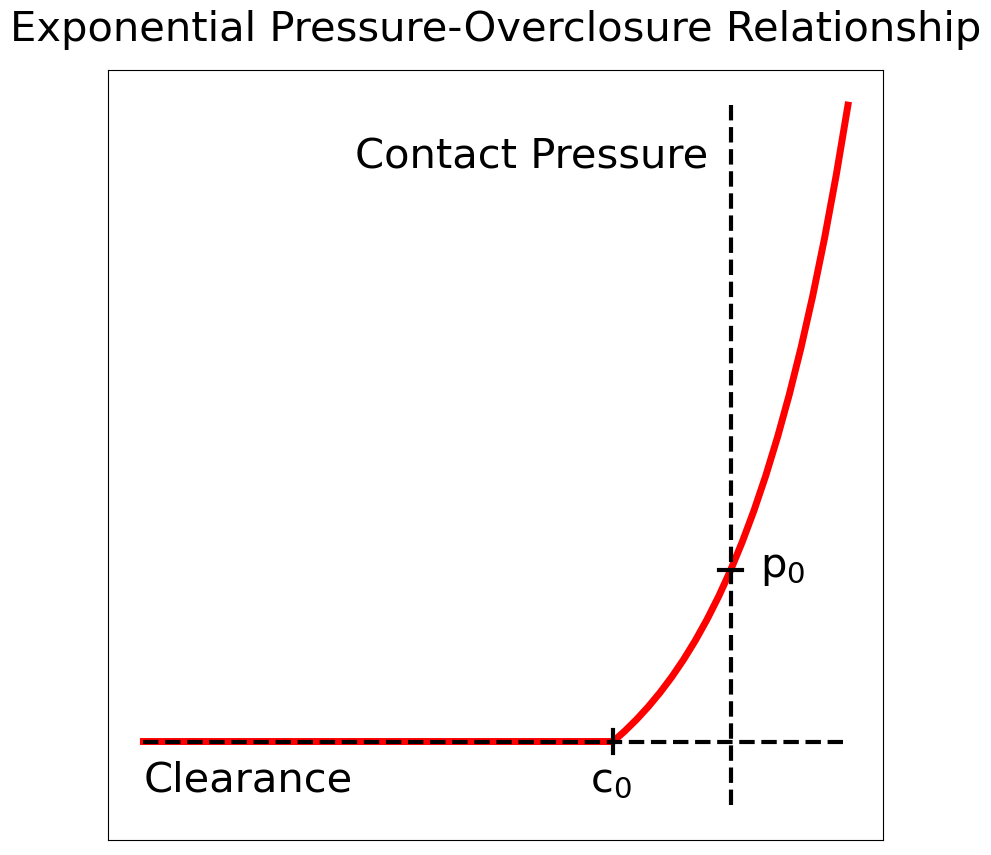}
	\caption{Softened Contact with Exponential Law}
	\label{Fig:abaqus conract exponential}
\end{figure}
A softened contact with an exponential law is employed \cite{abaqus_manual} in this work and given in \cref{Fig:abaqus conract exponential}. In this contact formulation, the surfaces begin to transfer contact pressure once the clearance between them, measured in the contact (normal) direction, decreases to c$_0$. By randomly sampling the clearance values between 2 and 5 $\mu$m in all contact definitions, many interference conditions were created while avoiding costly perturbations in geometry or remeshing.
\par Four clearances representing the interferences between the barrel and the first three lenses and the mutual interference between the first two lenses defined the input features for the neural network. The targets for the neural network model are deformed coordinates of the characteristic points of lens surfaces calculated by the finite element analysis. Surface displacement data is used to estimate the errors in positioning components in the assembly and perform Zernike fitting and calculate the corresponding optical performance responses. A few thousand data samples are generated using high-throughput computing capabilities of several nodes of a high-performance computing (HPC) cluster. In addition, parallel computing capabilities of the FEA code \cite{abaqus_manual} helped to reduce run time on each computing node. Approximately 80\% of the generated data samples are randomly selected for training, while the remaining 20\% is set aside for testing.

\subsection{Introduction to Dense Neural Networks} \label{Sec:DNN introduction}
\begin{figure}[H]
	\centering
	\includegraphics[width=0.6\textwidth]{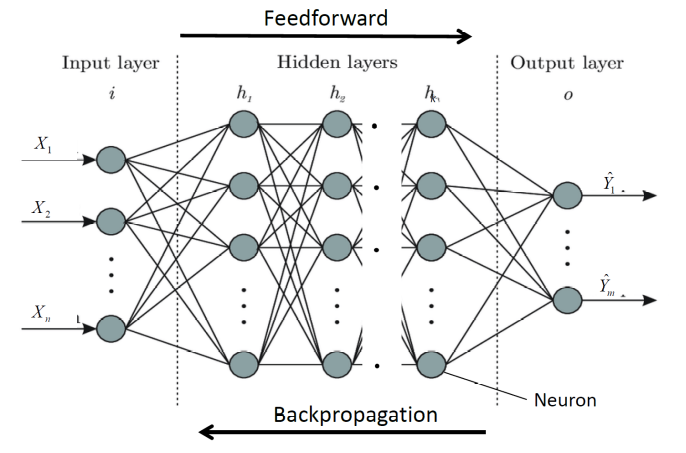}
	\caption{Feedforward Dense Neural Network}
	\label{Fig:DNN schematic}
\end{figure}

Deep learning is a subcategory of machine learning which is inspired by the configuration and functionality of a brain. Deep learning models are made of neural networks. Neural networks are composed of layers of interconnected, individual unit cells, named neurons, joined to other neurons’ layers. \Cref{Fig:DNN schematic} illustrates the feedforward dense neural network used in this work, consisting of linked layers of neurons that calculate the vector output predictions $\hat{\bm{Y}}$ based on input vector data $\bm{X}$.

After receiving input, the layers of neurons transmit information forward to the next layers, and this forms a network that learns with some feedback process. The layers in between input and output layers are called hidden layers, and the number of hidden layers represents a neural network’s deepness. Neurons of successive layers are connected through an accompanying weights and biases, marked $\bm{W}$ and $\bm{b}$ respectively. For a layer $l$, the predicted output $\hat{\bm{O}}^{[l]}$ is calculated as:

\begin{equation}
\begin{split}
\bm{Z}^{[l]} & = \bm{W}^{[l]} \hat{\bm{O}}^{[l-1]} + \bm{b}^{[l]} \\
\hat{\bm{O}}^{[l]} & = f^{[l]}(\bm{Z}^{[l]})
\end{split}
\label{Eq:DNN feedforward}
\end{equation}

where $\bm{W}^{[l]}(n_l \times n_{l-1})$ is a matrix of weights and $\bm{b}^{[l]} ( n_{l-1} \times 1)$ is a vector of biases, which are updated after every training pass. The $\bm{Z}$ vector, calculated from weights and biases, is further transformed by an activation function $f^{[l]}$ into output for every neuron in the layer $l$. The activation functions in neural networks are nonlinear functions such as Hyperbolic Tangent, Sigmoid and Rectified Linear Unit (ReLu). They allow the neural network to learn nearly any complicated functional relation between inputs and outputs. At the end of the each feed-forward pass, the loss function $\mathcal{L}$ calculates a loss value that shows how well the network's predictions $\hat{\bm{Y}}$ compare with targets $\bm{Y}$. One such commonly used loss function, called the mean squared error (MSE) is given in \cref{Eq:DNN loss func}.

\begin{equation}
\mathcal{L}(\bm{Y},\hat{\bm{Y}}) = \frac{1}{m} \sum_{i=1}^{m}(y_i - \hat{y_i})^2
\label{Eq:DNN loss func}
\end{equation}
where, $m$ is the size of the sample set. Then, in a so-called backpropagation procedure, the optimizer minimizes loss value iteratively with some optimization techniques such as gradient descent in \cref{Eq:DNN backprop}. The learning rate $\gamma$ is an important hyper parameter which controls how much the weights and biases of our network are adjusted with respect to the loss gradient in the learning process. The last layer's gradients of loss function $\mathcal{L}$ with respect to the weights are calculated first, and the weights are updated for each of its nodes. Using the chain rule of derivatives, the gradients at the previous layer are calculated and the weights are updated, and the same procedure is repeated backward up until all of the layers have had their weights updated \cite{pattanayak2017pro}. Then, a new forward propagation iteration $k+1$ starts again. After a sufficient number of feedforward and backpropagation iterations, the series $\bm{W}^k$ should converge toward a minima of loss function. The same backpropagation pattern is used for updating the biases.

\begin{equation}
\begin{split}
W_{ij}^{k+1} & = W_{ij}^{k} - \gamma \frac{\partial \mathcal{L}}{\partial W_{ij}^{k}} \\
b_i^{k+1} & = b_i^k - \gamma \frac{\partial \mathcal{L}}{\partial b_i^k}
\end{split}
\label{Eq:DNN backprop}
\end{equation}

\subsection{Sensitivity and Uncertainty Analyses}\label{Sec:sensitivity description}

Sensitivity analysis is used to assess the impact of the perturbation in an input on an output. Let $f$ denote the model function that maps an input vector $\bm{X} = [X_1,X_2,\dots X_d]$ to a scalar output $Y$; thus, $Y=f(\bm{X})$. In this case, for example, the deformations of each lens surface are functions of the interference values. Partial derivative of the output $Y$ with respect to a particular input $X_i$ can be used to define the sensitivity of the $Y$ to $X_i$. The partial derivative has to be evaluated at a particular value of the input: $\bm{X}=\bm{\hat{X}}$. This method estimates the local sensitivity at $\bm{\hat{X}}$. However, for most practical problems, the relationship between the inputs and outputs is highly nonlinear. Thus, the partial derivatives vary significantly from one design point to another. Evaluation of the local sensitivity at multiple design points gives massive data which makes the analysis difficult. Moreover, it does not provide a holistic perspective of the sensitivity. Hence, we use the global sensitivity analysis in this work.

We define the global sensitivity using a variance based analysis, also known as the Sobol method \cite{sobol2001global}. The relation $Y=f(\bm{X})$ is expanded as follows:
\begin{equation}
Y = f(\bm{X}) = f_0 + \sum_{i=1}^{d} f_i(X_i) + \sum_{i<j}^{d} f_{i,j}(X_i,X_j) + \dots + f_{1,2,\dots,d}(X_1,X_2,\dots X_d)
\label{Eq:sobol decomp}
\end{equation}
where, each term of the summation is a function over a subset of inputs. For instance, $f_i$ is a function of a single  component of the input vector $X_i$, $f_{i,j}$ is a function of two components $X_i$ and $X_j$ and so on. For a $d$ dimensional input space, there are $2^d$ in the summation. If each of the above functions has zero mean, this decomposition is known as ANOVA (analysis of variances):
\begin{equation}
\int f_{i_1,i_2,\dots,i_s} (X_{i_1}, X_{i_2}, \dots, X_{i_s}) dX_k = 0 \hspace{0.2cm} \text{for} \hspace{0.2cm} k=i_1,i_2,\dots,i_s
\end{equation}
If the above condition is satisfied, it can be shown that the functions are orthogonal and thus, the decomposition in \cref{Eq:sobol decomp} is unique \cite{saltelli2008global}. For a square-integrable function $f(\bm{X})$, squaring and integrating \cref{Eq:sobol decomp} gives:
\begin{equation}
\int Y^2 d\bm{X}  - f_0^2 = \sum_{s=1}^{d} \sum_{i_1<\dots<i_s}^{d} \int f^2_{i_1,\dots,i_s} d X_{i_1} \dots X_{i_s}
\label{Eq:sobol decomp integrate}
\end{equation}
Due to the orthogonality, the cross terms such as $\int f_{i_1}f_{i_2} dX_{i_1}dX_{i_2} \forall i_1 \neq i_2$ are zero. The left hand side of \cref{Eq:sobol decomp integrate} is the total variance in the output $Y$ and the right hand side is the summation of variances due to various subsets of the inputs. Therefore, the variance in $Y$ can be decomposed into variances caused by individual inputs and their interactions:
\begin{equation}
Var(Y) = \sum_{i=1}^{d} V_i + \sum_{i<j}^{d} V_{i,j} + \dots + V_{1,2,\dots,d}
\label{Eq:sobol decomp variance}
\end{equation}
The ratio of individual variance terms in \cref{Eq:sobol decomp variance} to the total variance is defined as sensitivity index. Dividing by the total variance gives:
\begin{equation}
1 = \sum_{i=1}^{d} S_i + \sum_{i<j}^{d} S_{i,j} + \dots + S_{1,2,\dots,d}
\label{Eq:sobol decomp sens indices}
\end{equation}
For instance, $S_i=V_i/Var(Y)$ and $S_{i,j}=V_{i,j}/Var(Y)$. Other higher order indices are similarly defined. Thus, all these $2^d-1$ indices sum to unity and are non-negative. We define the term $S_{T_i}$ corresponding each input $X_i$ as the sum of all the individual $2^{d-1}$ indices with the $i^{\text{th}}$ term present. for instance, for a three dimensional input space ($d=3$), the total Sobol index for the first input is given by $S_{T_1} = S_1 + S_{1,2} + S_{1,3} + S_{1,2,3}$. $S_{T_i}$ signifies the total contribution of the $i^{\text{th}}$ input in the variance of the output. However, the sum $\sum_{i=1}^{d} S_{T_i}$ is typically greater than unity since the terms with multiple inputs are counted more than once. In this work, we present the total Sobol indices of each output (deformations at various locations of the lens surfaces) with respect ot each input (interferences between the lenses). For simple functions, we can evaluate the integrals in \cref{Eq:sobol decomp integrate} analytically. However, for practical problems, the Monte-Carlo method is used to numerically estimate the Sobol indices. Brute force calculation is $\mathcal O (N^2)$ where, $N$ is the number of Monte-Carlo samples \cite{saltelli2008global}. Since the convergence rate of Monte-Carlo algorithm is $\mathcal O (N^{-1/2})$ \cite{caflisch1998monte}, the sample size $N$ can of the order of $10^5\sim10^6$. These computations are fairly expensive even with the use of surrogate models such as neural networks. \citet{saltelli2008global} proposed an algorithm which requires $\mathcal O (N(d+2))$ computations.

\section{Results and Discussions}

\subsection{Deep Neural Network Training and Testing}
Training of neural network requires multiple hyper-parameters such as number of hidden layers and neurons, activation function, learning rate, dropout factor etc. These hyper-parameters are fine tuned by randomly splitting the data into two subsets: training and validation. The training set is used for the back-propagation algorithm described in \cref{Sec:DNN introduction}. The loss function is evaluated on the validation set and compared with the training loss. A shallow network which has fewer hidden layers and neurons gives higher error in fitting the training set. This is known as under-fitting or bias. Depth of the network is increased by adding more hidden layers and neurons. Such a network with higher nonlinearity improves the prediction accuracy on the training set. Excessively deep networks can fit the training set with high accuracy but fail to fit the unseen validation set. This phenomenon is known as over-fitting or variance. It is important to have a network with low bias and low variance which can fit the training data successfully as well as generalize on the validation data. Such a well trained network is further tested on an unseen test data set. \Cref{Tab:DNN hyperparameters} lists the values of all the hyper-parameters used in this work.

\begin{table}[H]
	\centering
	\begin{tabular}{|c|c|}
		\hline
		Hyper-Parameters                & Values             \\ \hline
		Size of Training Set            & 2500               \\ \hline
		Size of Testing Set             & 300                \\ \hline
		Validation Split                & 10\%               \\ \hline
		No. of Hidden Layers            & 10                 \\ \hline
		No. of Neurons per Hidden Layer & 200                \\ \hline
		No. of Trainable Parameters     & 443984             \\ \hline
		Learning Rate                   & 0.01               \\ \hline
		Dropout Factor                  & 0.1                \\ \hline
		No. of Epochs                   & 500                \\ \hline
		Loss Function                   & Mean Squared Error \\ \hline
		Hidden Layers Activation        & ReLU               \\ \hline
		Output Layer Activation         & Linear             \\ \hline
		Optimization Algorithm          & Adam \cite{kingma2014adam}              \\ \hline
	\end{tabular}
	\caption{Hyper-Parameters of the Deep Neural Network}
	\label{Tab:DNN hyperparameters}
\end{table}
We have estimated the prediction accuracy of the neural network using the coefficient of determination \cite{cameron1997r}:
\begin{equation}
\text{Accuracy: } R^2 = 1 - \frac{\sum_{i=1}^{m} (y_i - \hat{y_i})^2}{\sum_{i=1}^{m} (y_i - \text{mean}(\bm{Y}))^2}
\label{Eq:accuracy}
\end{equation}
where, $\bm{Y}=[y_i]$, $1 \leq i \leq m$ is the target data set obtained from the numerical simulations, $\hat{y_i}$ is the corresponding predicted set by the neural network and $m$ is the sample size. Similarly, percentage errors are defined as:
\begin{equation}
\text{Average percent error: } 100 \times \frac{1}{m} \frac{\sum_{i=1}^{m} ||y_i - \hat{y_i}||}{\max_{i=1}^m ||y_i||}
\label{Eq:average percent error}
\end{equation}
\begin{equation}
\text{Maximum percent error: } 100 \times \frac{\max_{i=1}^{m} ||y_i -\hat{y_i}||}{\max_{i=1}^m ||y_i||}
\label{Eq:maximum percent error}
\end{equation}

\begin{figure}[H]
	\centering
	\begin{subfigure}[t]{0.45\textwidth}
		\includegraphics[width=\textwidth]{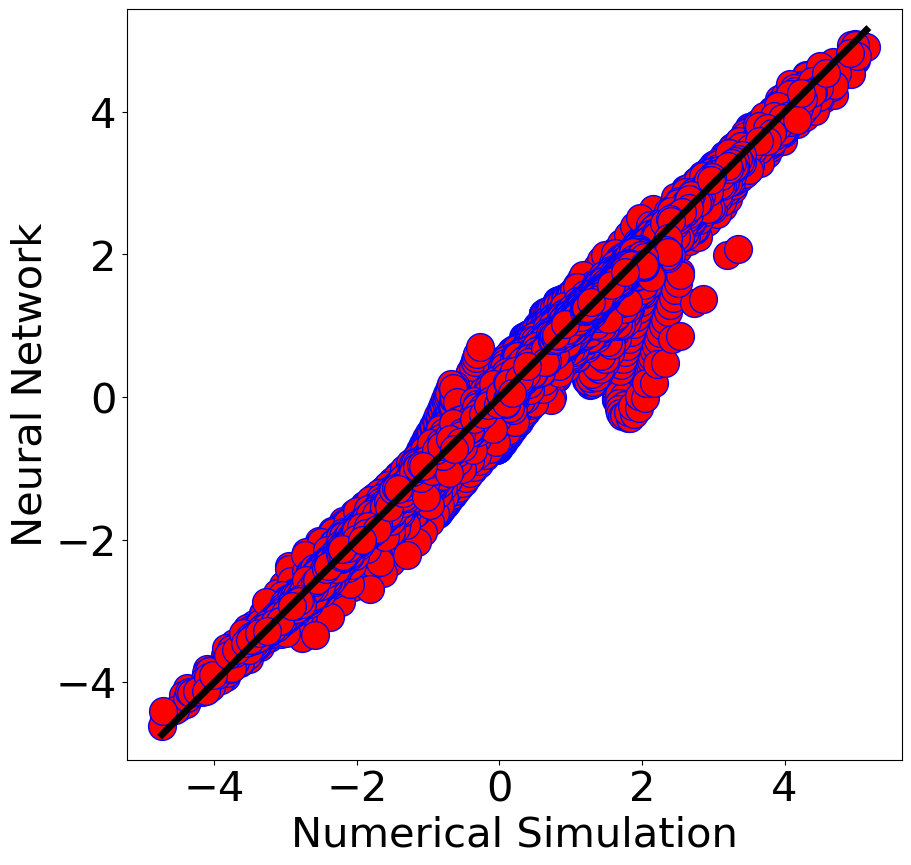}
		\caption{Training Set}
	\end{subfigure}
	\hspace{0.05\textwidth}
	\begin{subfigure}[t]{0.45\textwidth}
		\includegraphics[width=\textwidth]{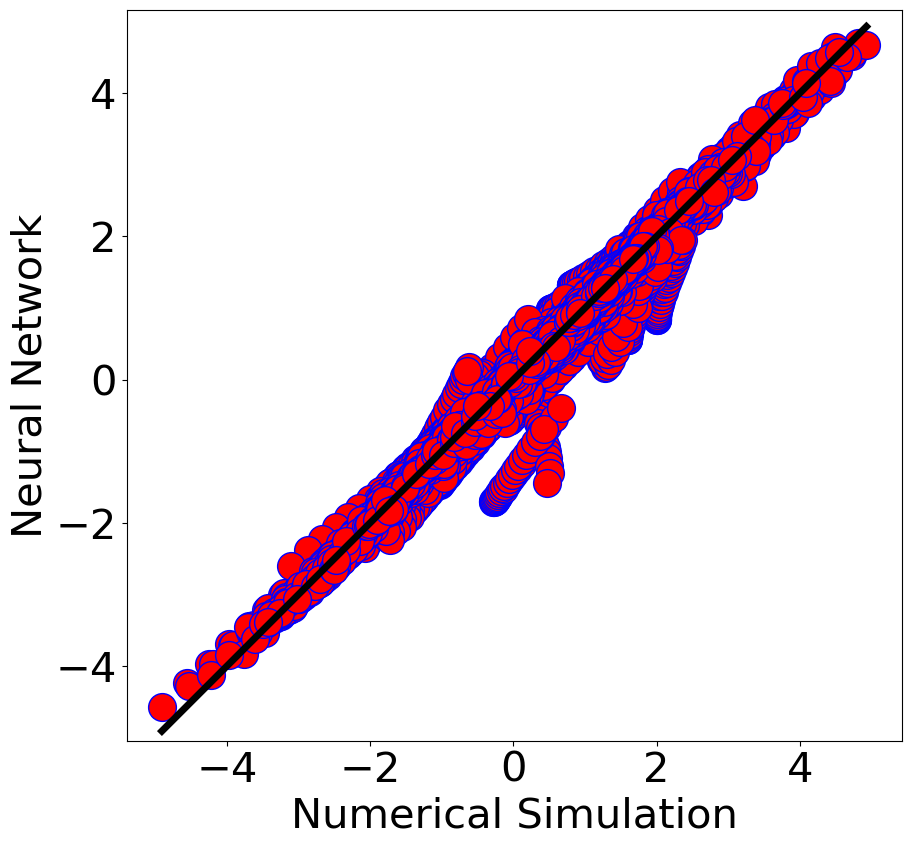}
		\caption{Testing Set}
	\end{subfigure}
	\caption{Comparison of Numerical Simulations and Neural Network Predictions}
	\label{Fig:DNN comparison}
\end{figure}

\begin{table}[H]
	\centering
	\begin{tabular}{|c|c|c|}
		\hline
		& Training Set & Testing Set \\ \hline
		Accuracy ($R^2$)      & 0.994368     & 0.992977    \\ \hline
		Average Percent Error & 0.948018\%     & 1.04884\%     \\ \hline
		Maximum Percent Error & 41.3737\%     & 38.7922\%    \\ \hline
	\end{tabular}
	\caption{Accuracy and Error of the Deep Neural Network}
	\label{Tab:DNN accuracy and error}
\end{table}
The accuracies and errors for the training and testing sets are listed in \cref{Tab:DNN accuracy and error}. If a model is perfect and fits the data exactly, the coefficient of determination ($R^2$) takes a value of unity \cite{cameron1997r}. However, for practical models, the $R^2$ is found to be less than unity. Hence, a value close to unity is indicative of high accuracy. In this case, we see that the accuracy for both the sets is higher than 0.99 and the average error is around 1\%. Note that accurate networks may have a few outliers which manifest in the maximum error. Low error and high accuracy for the training set shows that the network has less bias. Moreover, similar errors and accuracies for both the data sets indicate that the chosen hyper-parameters give less variance. In order to get a visual understanding of the network's accuracy, \cref{Fig:DNN comparison} plots the estimate from the neural network versus the `ground truth' which is the numerical prediction in this case. Both the axes are non-dimensionalized by subtracting the mean and dividing by the standard deviation of the numerical simulations. We can observe that most the points follow the trend line $Y=X$. The outliers mentioned above can be seen in this plot as the points away from the trend line.

\subsection{Sensitivity Analysis} \label{Sec:Sensitivity Analysis results}

\begin{figure}[H]
	\centering
	\begin{subfigure}[t]{0.45\textwidth}
		\includegraphics[width=\textwidth]{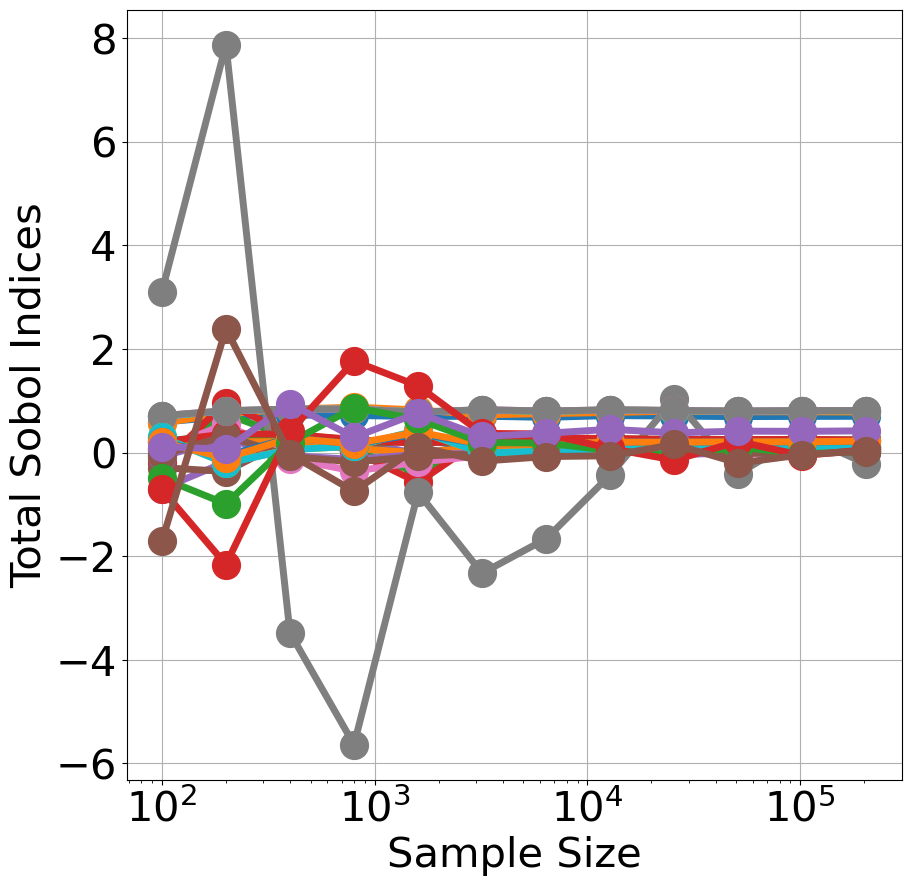}
		\caption{Interference between Barrel and Lens 1}
	\end{subfigure}
	\hspace{0.05\textwidth}
	\begin{subfigure}[t]{0.45\textwidth}
		\includegraphics[width=\textwidth]{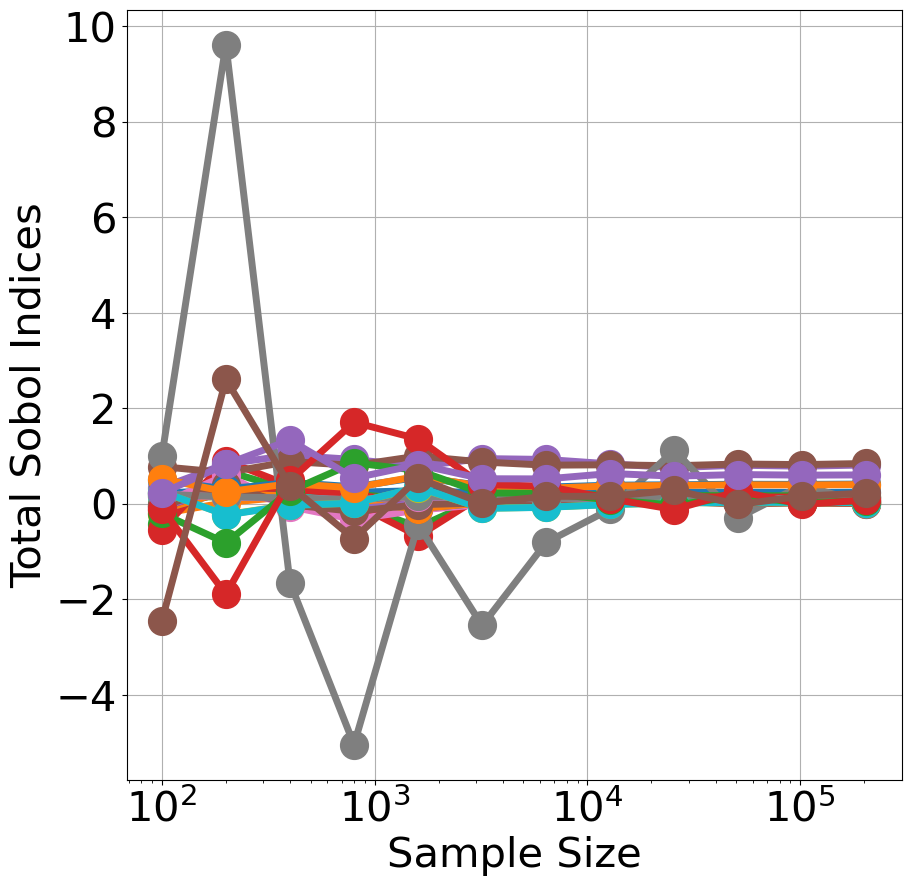}
		\caption{Interference between Barrel and Lens 3} \vspace{0.25cm}
	\end{subfigure}
	\begin{subfigure}[t]{0.45\textwidth}
		\includegraphics[width=\textwidth]{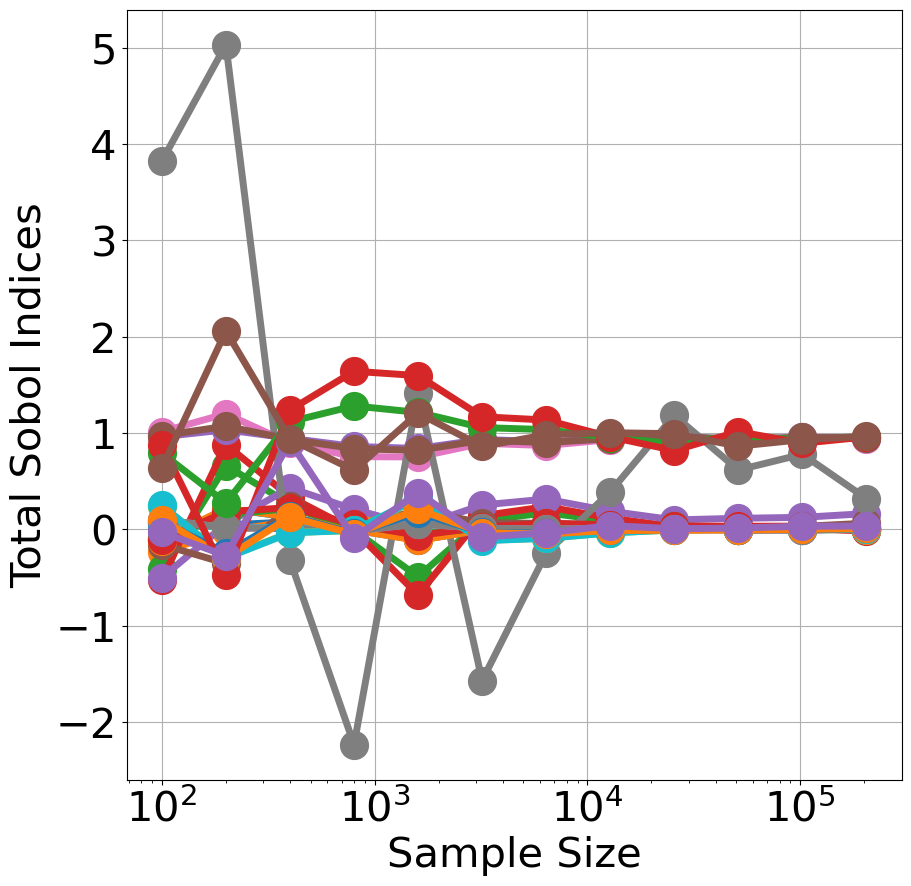}
		\caption{Interference between Barrel and Lens 4}
	\end{subfigure}
	\hspace{0.05\textwidth}
	\begin{subfigure}[t]{0.45\textwidth}
		\includegraphics[width=\textwidth]{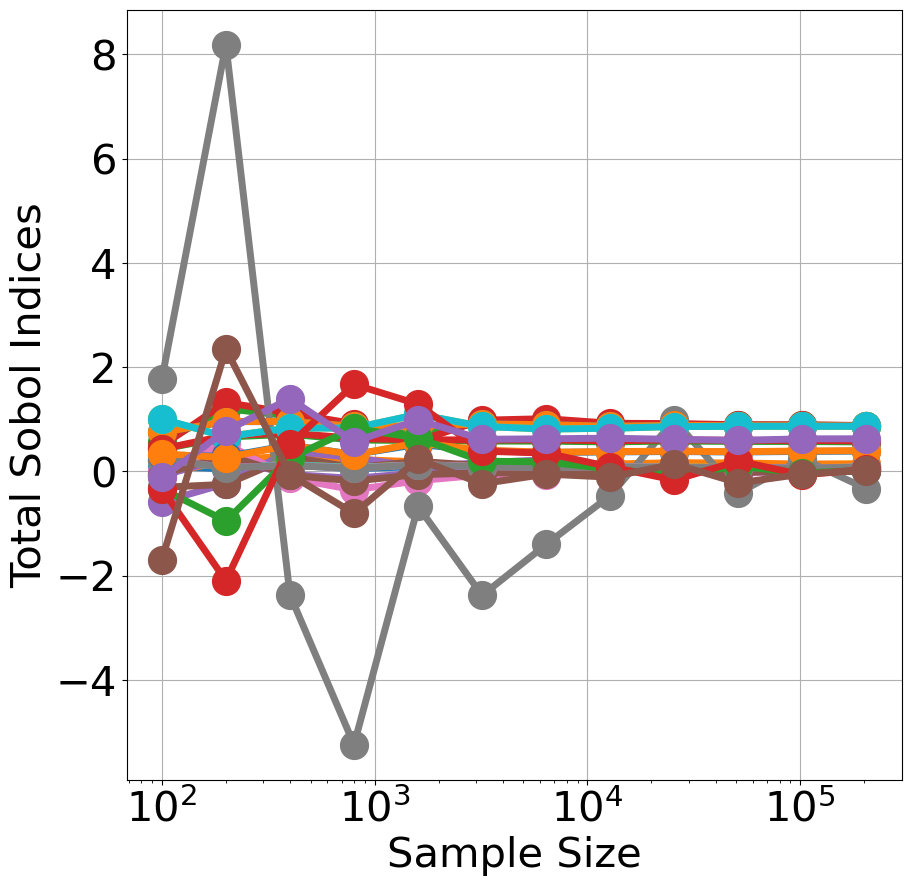}
		\caption{Interference between Lenses 1 and 2}
	\end{subfigure}
	\caption{Convergence of Sobol Indices for 24 Outputs (Deformations) and 4 Inputs (Interferences)}
	\label{Fig:sobol convergence}
\end{figure}
As described in \cref{Sec:sensitivity description}, the Monte-Carlo based algorithm proposed by \citet{saltelli2008global} is used to estimate the global sensitivity indices. For practical engineering problems, since the analytical solutions are unavailable, the sample size is increased till the asymptotic convergence of the solution. \Cref{Fig:sobol convergence} plots the convergence of the Sobol indices separately for each input. For 4 inputs and 26 outputs, there are a total of $4 \times 24 = 96$ indices. The sample size is increased exponentially by a factor of 2 from 100 to 2E5. We can observe that the although the initial estimates at lower sample sizes are inaccurate (sometimes even negative), asymptotically stationary solutions are obtained beyond the sample size of 1E4. Hence, the average of last three estimates is documented as the final Sobol index. Note that the sample sizes are of the order of 1E5 and thus, it is necessary to use the neural network for function evaluations. It is computationally expensive to use the high fidelity numerical computations.
\par \Cref{Tab:sobol indices} plots the total Sobol indices for 24 outputs and 4 inputs. The radial deformation ($\Delta r$) at the outer end (r max) of the lens and the axial deformation ($\Delta z$) at the center of the lens (r min) and its outer end (r max) are important outputs which affect the optical properties of the system. Hence, these 3 outputs are defined for both the surfaces of each of the four lenses. These outputs are grouped according to lens surfaces. Each output is highly sensitive to a single input out of the four interferences. Those higher Sobol indices are emphasized in the table. For instance, all the six deformations on lens 1 are most sensitive to the interference between the barrel and lens 1. They are also sensitive to other deformations since the later lenses slightly deform the barrel. Similarly, deformations on lens 2 are most sensitive to the interference between lenses 1 and 2. In the current setup, lens 2 does not directly come into contact with the barrel and hence, barrel-lens 2 interference is missing. We observe similar behavior for the third and fourth lenses.

\begin{table}[H] 
	\centering
	\resizebox{\textwidth}{!}{%
		\begin{tabular}{|c|c|c|cccc|}
			\hline
			\multirow{2}{*}{\begin{tabular}[c]{@{}c@{}}Lens \\ Number\end{tabular}} & \multirow{2}{*}{Surface} & \multirow{2}{*}{\begin{tabular}[c]{@{}c@{}}Outputs: \\ Deformations\end{tabular}} & \multicolumn{4}{c|}{Inputs: Interferences}                                                                                                                               \\ \cline{4-7}
			&                          &                                                                                   & \multicolumn{1}{c|}{Barrel-Lens 1}            & \multicolumn{1}{c|}{Barrel-Lens 3}            & \multicolumn{1}{c|}{Barrel-Lens 4}            & Lens 1-Lens 2            \\ \hline
			\multirow{6}{*}{Lens 1}                                                 & \multirow{3}{*}{Bottom}  & $\Delta r$(r max)                                                                 & \multicolumn{1}{c|}{\textit{\textbf{0.7000}}} & \multicolumn{1}{c|}{0.2417}                   & \multicolumn{1}{c|}{0.0099}                   & 0.0387                   \\ \cline{3-7}
			&                          & $\Delta z$(r min)                                                                 & \multicolumn{1}{c|}{\textit{\textbf{0.8085}}} & \multicolumn{1}{c|}{0.1070}                   & \multicolumn{1}{c|}{0.0000}                   & 0.0776                   \\ \cline{3-7}
			&                          & $\Delta z$(r max)                                                                 & \multicolumn{1}{c|}{\textit{\textbf{0.8105}}} & \multicolumn{1}{c|}{0.1083}                   & \multicolumn{1}{c|}{0.0000}                   & 0.0754                   \\ \cline{2-7}
			& \multirow{3}{*}{Top}     & $\Delta r$(r max)                                                                 & \multicolumn{1}{c|}{\textit{\textbf{0.7854}}} & \multicolumn{1}{c|}{0.0503}                   & \multicolumn{1}{c|}{0.0000}                   & 0.1428                   \\ \cline{3-7}
			&                          & $\Delta z$(r min)                                                                 & \multicolumn{1}{c|}{\textit{\textbf{0.8073}}} & \multicolumn{1}{c|}{0.1128}                   & \multicolumn{1}{c|}{0.0000}                   & 0.0730                   \\ \cline{3-7}
			&                          & $\Delta z$(r max)                                                                 & \multicolumn{1}{c|}{\textit{\textbf{0.8104}}} & \multicolumn{1}{c|}{0.1121}                   & \multicolumn{1}{c|}{0.0000}                   & 0.0718                   \\ \hline
			\multirow{6}{*}{Lens 2}                                                 & \multirow{3}{*}{Bottom}  & $\Delta r$(r max)                                                                 & \multicolumn{1}{c|}{0.0968}                   & \multicolumn{1}{c|}{0.0154}                   & \multicolumn{1}{c|}{0.0000}                   & \textit{\textbf{0.8814}} \\ \cline{3-7}
			&                          & $\Delta z$(r min)                                                                 & \multicolumn{1}{c|}{0.1092}                   & \multicolumn{1}{c|}{0.0229}                   & \multicolumn{1}{c|}{0.0009}                   & \textit{\textbf{0.8703}} \\ \cline{3-7}
			&                          & $\Delta z$(r max)                                                                 & \multicolumn{1}{c|}{0.1200}                   & \multicolumn{1}{c|}{0.0289}                   & \multicolumn{1}{c|}{0.0055}                   & \textit{\textbf{0.8630}} \\ \cline{2-7}
			& \multirow{3}{*}{Top}     & $\Delta r$(r max)                                                                 & \multicolumn{1}{c|}{0.0887}                   & \multicolumn{1}{c|}{0.0153}                   & \multicolumn{1}{c|}{0.0000}                   & \textit{\textbf{0.8854}} \\ \cline{3-7}
			&                          & $\Delta z$(r min)                                                                 & \multicolumn{1}{c|}{0.1090}                   & \multicolumn{1}{c|}{0.0227}                   & \multicolumn{1}{c|}{0.0009}                   & \textit{\textbf{0.8705}} \\ \cline{3-7}
			&                          & $\Delta z$(r max)                                                                 & \multicolumn{1}{c|}{0.1200}                   & \multicolumn{1}{c|}{0.0287}                   & \multicolumn{1}{c|}{0.0055}                   & \textit{\textbf{0.8634}} \\ \hline
			\multirow{6}{*}{Lens 3}                                                 & \multirow{3}{*}{Bottom}  & $\Delta r$(r max)                                                                 & \multicolumn{1}{c|}{0.0073}                   & \multicolumn{1}{c|}{\textit{\textbf{0.8011}}} & \multicolumn{1}{c|}{0.1437}                   & 0.0628                   \\ \cline{3-7}
			&                          & $\Delta z$(r min)                                                                 & \multicolumn{1}{c|}{0.2465}                   & \multicolumn{1}{c|}{\textit{\textbf{0.1653}}} & \multicolumn{1}{c|}{0.0221}                   & 0.5682                   \\ \cline{3-7}
			&                          & $\Delta z$(r max)                                                                 & \multicolumn{1}{c|}{0.2110}                   & \multicolumn{1}{c|}{\textit{\textbf{0.4107}}} & \multicolumn{1}{c|}{0.0042}                   & 0.3851                   \\ \cline{2-7}
			& \multirow{3}{*}{Top}     & $\Delta r$(r max)                                                                 & \multicolumn{1}{c|}{0.0786}                   & \multicolumn{1}{c|}{\textit{\textbf{0.8337}}} & \multicolumn{1}{c|}{0.0466}                   & 0.0554                   \\ \cline{3-7}
			&                          & $\Delta z$(r min)                                                                 & \multicolumn{1}{c|}{0.2484}                   & \multicolumn{1}{c|}{\textit{\textbf{0.1380}}} & \multicolumn{1}{c|}{0.0269}                   & 0.5886                   \\ \cline{3-7}
			&                          & $\Delta z$(r max)                                                                 & \multicolumn{1}{c|}{0.2145}                   & \multicolumn{1}{c|}{\textit{\textbf{0.3996}}} & \multicolumn{1}{c|}{0.0050}                   & 0.3940                   \\ \hline
			\multirow{6}{*}{Lens 4}                                                 & \multirow{3}{*}{Bottom}  & $\Delta r$(r max)                                                                 & \multicolumn{1}{c|}{0.0091}                   & \multicolumn{1}{c|}{0.0606}                   & \multicolumn{1}{c|}{\textit{\textbf{0.9455}}} & 0.0063                   \\ \cline{3-7}
			&                          & $\Delta z$(r min)                                                                 & \multicolumn{1}{c|}{0.0067}                   & \multicolumn{1}{c|}{0.0525}                   & \multicolumn{1}{c|}{\textit{\textbf{0.9514}}} & 0.0038                   \\ \cline{3-7}
			&                          & $\Delta z$(r max)                                                                 & \multicolumn{1}{c|}{0.0008}                   & \multicolumn{1}{c|}{0.0451}                   & \multicolumn{1}{c|}{\textit{\textbf{0.9376}}} & 0.0000                   \\ \cline{2-7}
			& \multirow{3}{*}{Top}     & $\Delta r$(r max)                                                                 & \multicolumn{1}{c|}{0.0578}                   & \multicolumn{1}{c|}{0.1975}                   & \multicolumn{1}{c|}{\textit{\textbf{0.6492}}} & 0.1107                   \\ \cline{3-7}
			&                          & $\Delta z$(r min)                                                                 & \multicolumn{1}{c|}{0.0072}                   & \multicolumn{1}{c|}{0.0539}                   & \multicolumn{1}{c|}{\textit{\textbf{0.9504}}} & 0.0045                   \\ \cline{3-7}
			&                          & $\Delta z$(r max)                                                                 & \multicolumn{1}{c|}{0.0000}                   & \multicolumn{1}{c|}{0.0355}                   & \multicolumn{1}{c|}{\textit{\textbf{0.9265}}} & 0.0000                   \\ \hline
		\end{tabular}%
	}
	\caption{Total Sobol Indices}
	\label{Tab:sobol indices}
\end{table}

\subsection{Uncertainty Propagation Analysis}
\begin{figure}[H]
	\centering
	\begin{subfigure}[t]{0.45\textwidth}
		\includegraphics[width=\textwidth]{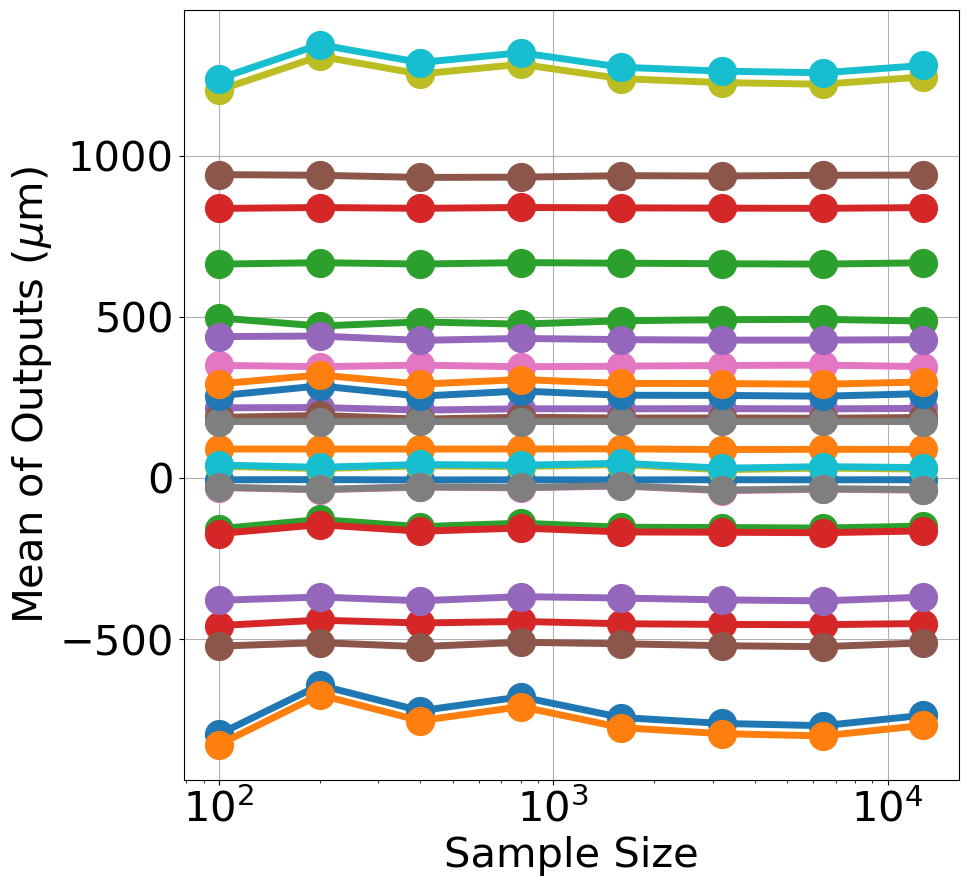}
		\caption{Mean ($\mu$m)}
	\end{subfigure}
	\hspace{0.05\textwidth}
	\begin{subfigure}[t]{0.45\textwidth}
		\includegraphics[width=\textwidth]{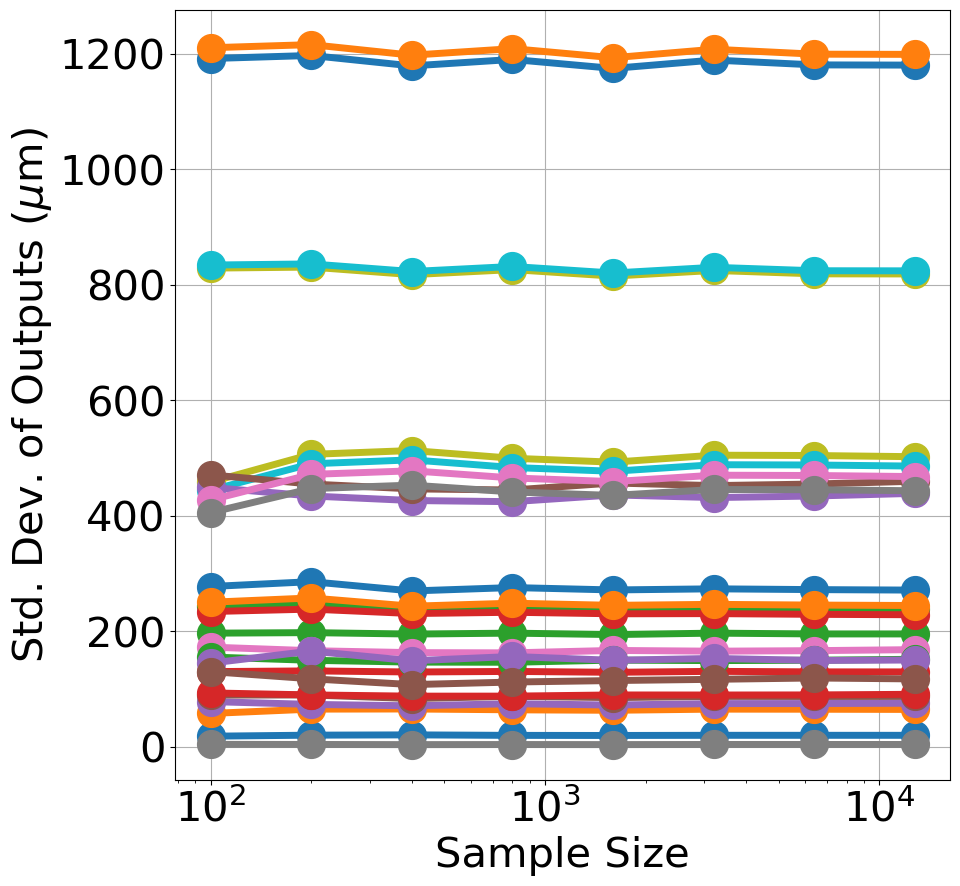}
		\caption{Standard Deviation ($\mu$m)}
	\end{subfigure}
	\caption{Convergence of the Monte-Carlo Algorithm for 24 Outputs}
	\label{Fig:UQ mean std convergence}
\end{figure}

The following four interference values are assumed to independently follow uniform distributions:
\begin{enumerate}
	\item Barrel-Lens 1 $\sim \mathcal{U}(2.77316, 4.32155)$ $\mu$m
	\item Barrel-Lens 3 $\sim \mathcal{U}(3.02553, 4.79024)$ $\mu$m
	\item Barrel-Lens 4 $\sim \mathcal{U}(2.23457, 4.70370)$ $\mu$m
	\item Lens 1-Lens 2 $\sim \mathcal{U}(2.82143, 4.96429)$ $\mu$m
\end{enumerate}
In this work, we have modeled the propagation of uncertainty in the above four input parameters on the 24 deformations described in \cref{Sec:Sensitivity Analysis results}. \Cref{Fig:UQ mean std convergence} plots the convergence of the Monte-Carlo algorithm for the prediction of the means and standard deviations of the outputs. It can be seen that the convergence is achieved beyond 3200 samples. We have used a sample size of 12800 for all the computations in this section. This shows the computational benefit of using the neural network as a surrogate model instead of complete numerical simulations.

\begin{table}[H]
	\centering
	\resizebox{\textwidth}{!}{%
		\begin{tabular}{|c|c|c|c|c|}
			\hline
			Lens Number             & Surface                 & Outputs: Deformations & Mean ($\mu$m) & Standard Deviation ($\mu$m) \\ \hline
			\multirow{6}{*}{Lens 1} & \multirow{3}{*}{Bottom} & $\Delta r$(r max)     & -5.02         & 20.15                       \\ \cline{3-5}
			&                         & $\Delta z$(r min)     & 28.09         & 502.55                      \\ \cline{3-5}
			&                         & $\Delta z$(r max)     & -37.39        & 468.13                      \\ \cline{2-5}
			& \multirow{3}{*}{Top}    & $\Delta r$(r max)     & 89.12         & 64.95                       \\ \cline{3-5}
			&                         & $\Delta z$(r min)     & 32.57         & 486.28                      \\ \cline{3-5}
			&                         & $\Delta z$(r max)     & -35.59        & 443.61                      \\ \hline
			\multirow{6}{*}{Lens 2} & \multirow{3}{*}{Bottom} & $\Delta r$(r max)     & 487.34        & 195.68                      \\ \cline{3-5}
			&                         & $\Delta z$(r min)     & -735.74       & 1180.59                     \\ \cline{3-5}
			&                         & $\Delta z$(r max)     & 1244.88       & 819.00                      \\ \cline{2-5}
			& \multirow{3}{*}{Top}    & $\Delta r$(r max)     & -451.20       & 129.93                      \\ \cline{3-5}
			&                         & $\Delta z$(r min)     & -767.07       & 1199.16                     \\ \cline{3-5}
			&                         & $\Delta z$(r max)     & 1280.49       & 824.38                      \\ \hline
			\multirow{6}{*}{Lens 3} & \multirow{3}{*}{Bottom} & $\Delta r$(r max)     & 216.38        & 75.69                       \\ \cline{3-5}
			&                         & $\Delta z$(r min)     & -148.88       & 238.74                      \\ \cline{3-5}
			&                         & $\Delta z$(r max)     & 262.14        & 271.58                      \\ \cline{2-5}
			& \multirow{3}{*}{Top}    & $\Delta r$(r max)     & 188.20        & 87.52                       \\ \cline{3-5}
			&                         & $\Delta z$(r min)     & -163.74       & 228.90                      \\ \cline{3-5}
			&                         & $\Delta z$(r max)     & 298.41        & 244.89                      \\ \hline
			\multirow{6}{*}{Lens 4} & \multirow{3}{*}{Bottom} & $\Delta r$(r max)     & 346.17        & 168.51                      \\ \cline{3-5}
			&                         & $\Delta z$(r min)     & -369.37       & 439.14                      \\ \cline{3-5}
			&                         & $\Delta z$(r max)     & 668.15        & 151.81                      \\ \cline{2-5}
			& \multirow{3}{*}{Top}    & $\Delta r$(r max)     & 175.65        & 4.29                        \\ \cline{3-5}
			&                         & $\Delta z$(r min)     & -511.03       & 460.12                      \\ \cline{3-5}
			&                         & $\Delta z$(r max)     & 839.51        & 91.18                       \\ \hline
		\end{tabular}%
	}
	\caption{Uncertainty Propagation: Means and Standard Deviations}
	\label{Tab:UQ mean and std}
\end{table}

The means and standard deviations for all 24 outputs are listed in \cref{Tab:UQ mean and std}. The input parameters (interferences) are assumed to vary uniformly within a range of a couple of microns. However, we observe that the standard deviations of most of the outputs is of the order of a few hundred microns. This shows the utility of the uncertainty propagation analysis. Highly nonlinear and complex models such as the ones described in this research typically magnifies the input uncertainty. Such an information is important in practice to estimate the possible ranges of the outputs. Note that these ranges cannot be computed using a few deterministic simulations.

\begin{figure}[H]
	\centering
	\begin{subfigure}[t]{0.45\textwidth}
		\includegraphics[width=\textwidth]{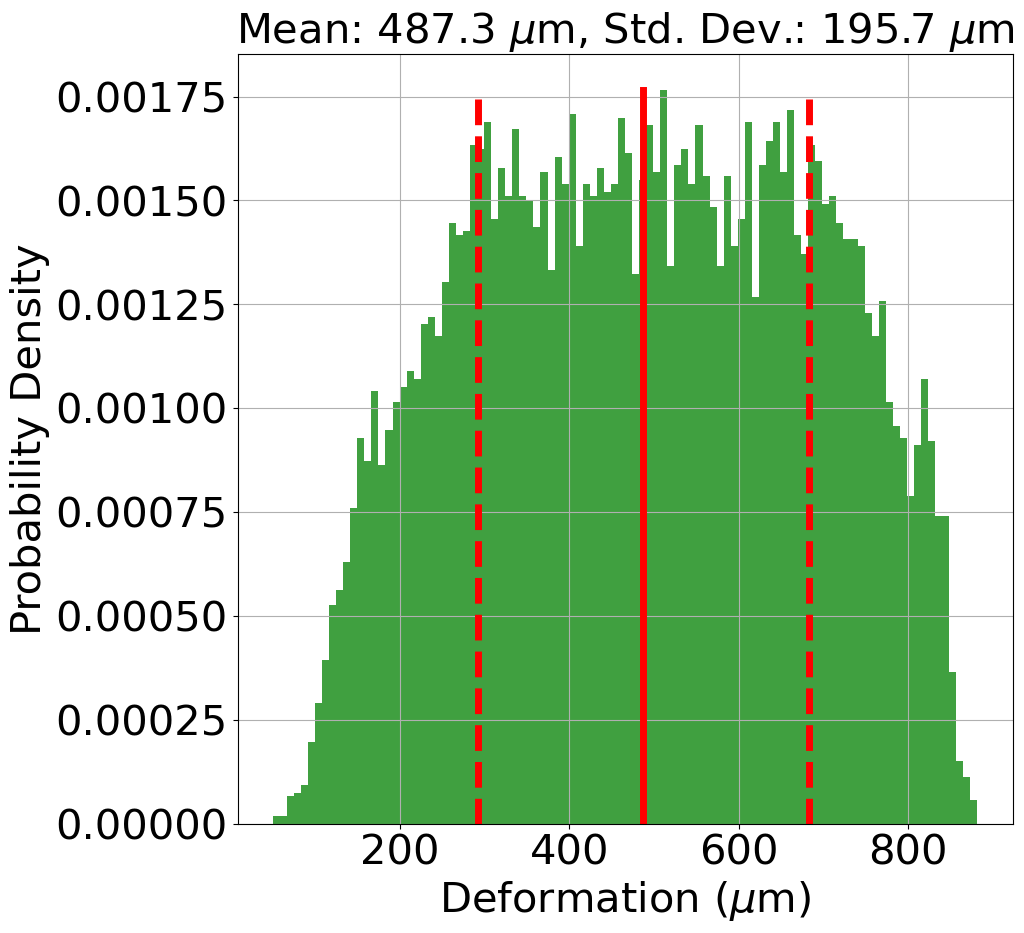}
		\caption{Lens 2 Bottom Surface}
	\end{subfigure}
	\hspace{0.05\textwidth}
	\begin{subfigure}[t]{0.45\textwidth}
		\includegraphics[width=\textwidth]{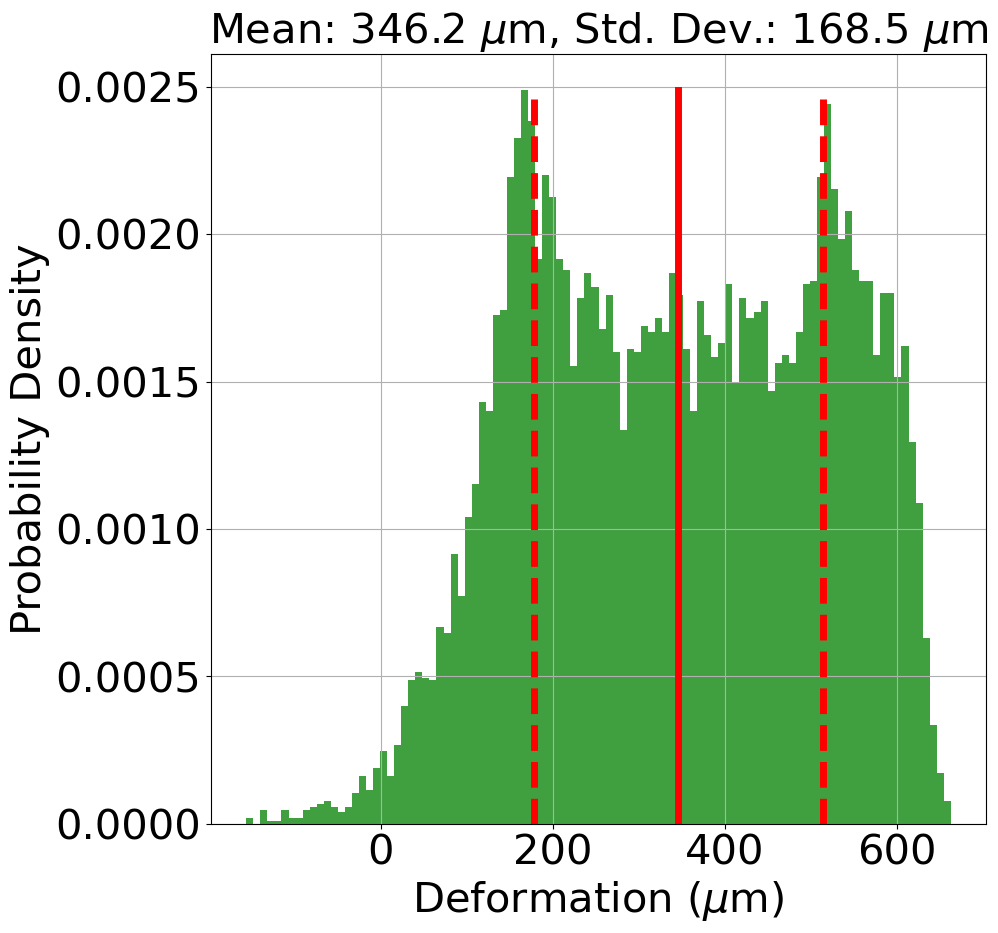}
		\caption{Lens 4 Bottom Surface}
	\end{subfigure}
	\caption{Histograms: $\Delta r$(r max)}
	\label{Fig:UQ pdf dr(r max)}
\end{figure}
\begin{figure}[H]
	\centering
	\begin{subfigure}[t]{0.45\textwidth}
		\includegraphics[width=\textwidth]{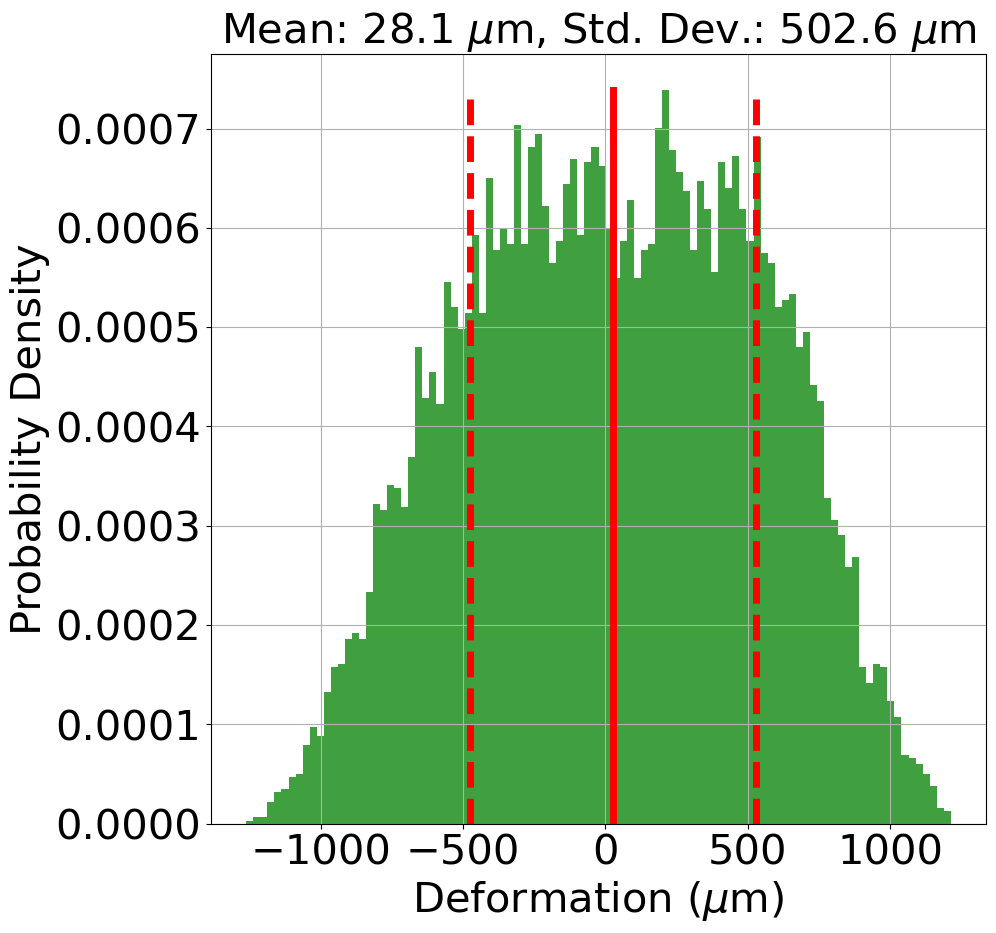}
		\caption{Lens 1 Bottom Surface}
	\end{subfigure}
	\hspace{0.05\textwidth}
	\begin{subfigure}[t]{0.45\textwidth}
		\includegraphics[width=\textwidth]{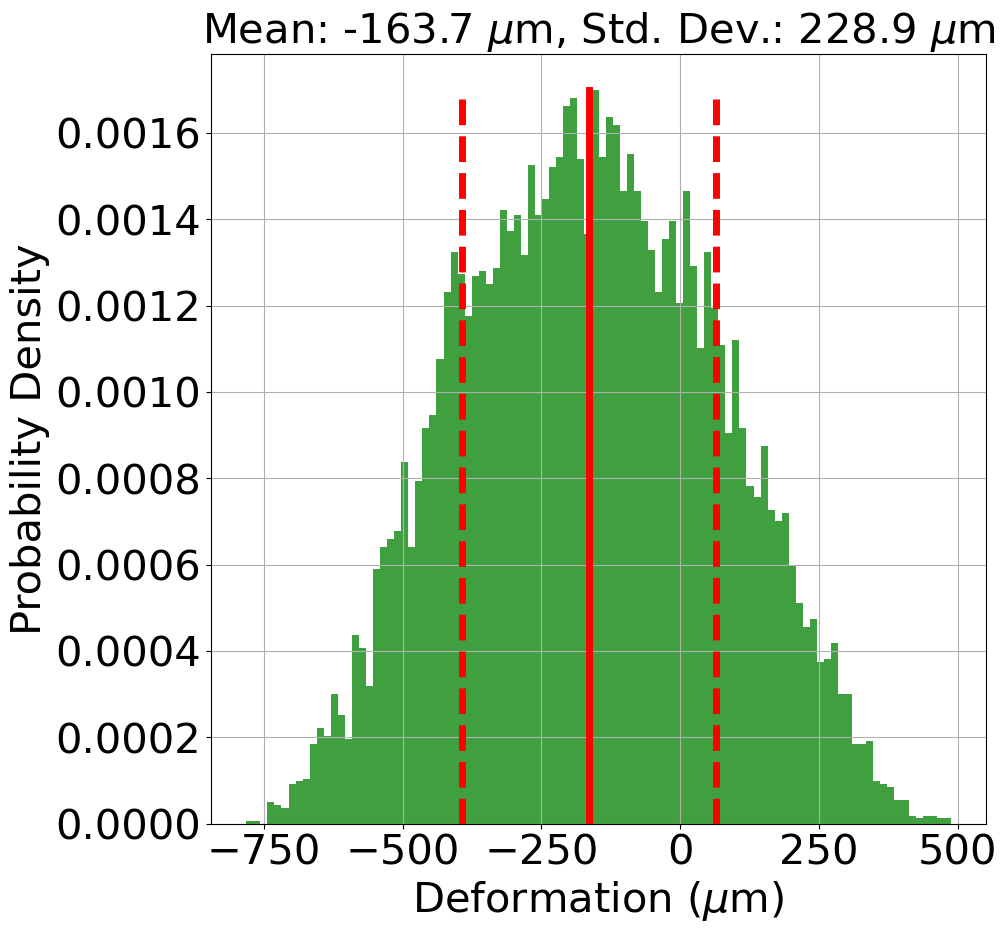}
		\caption{Lens 3 Top Surface}
	\end{subfigure}
	\caption{Histograms: $\Delta z$(r min)}
	\label{Fig:UQ pdf dz(r min)}
\end{figure}
\begin{figure}[H]
	\centering
	\begin{subfigure}[t]{0.45\textwidth}
		\includegraphics[width=\textwidth]{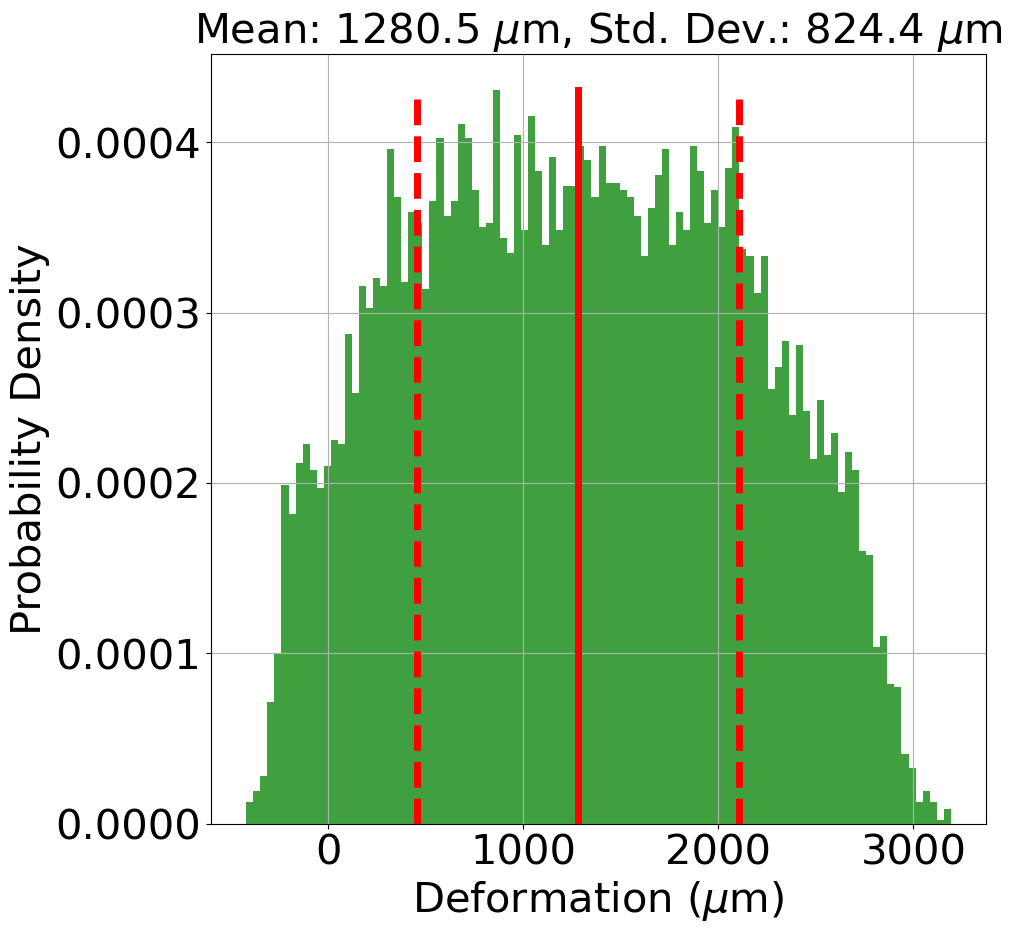}
		\caption{Lens 2 Top Surface}
	\end{subfigure}
	\hspace{0.05\textwidth}
	\begin{subfigure}[t]{0.45\textwidth}
		\includegraphics[width=\textwidth]{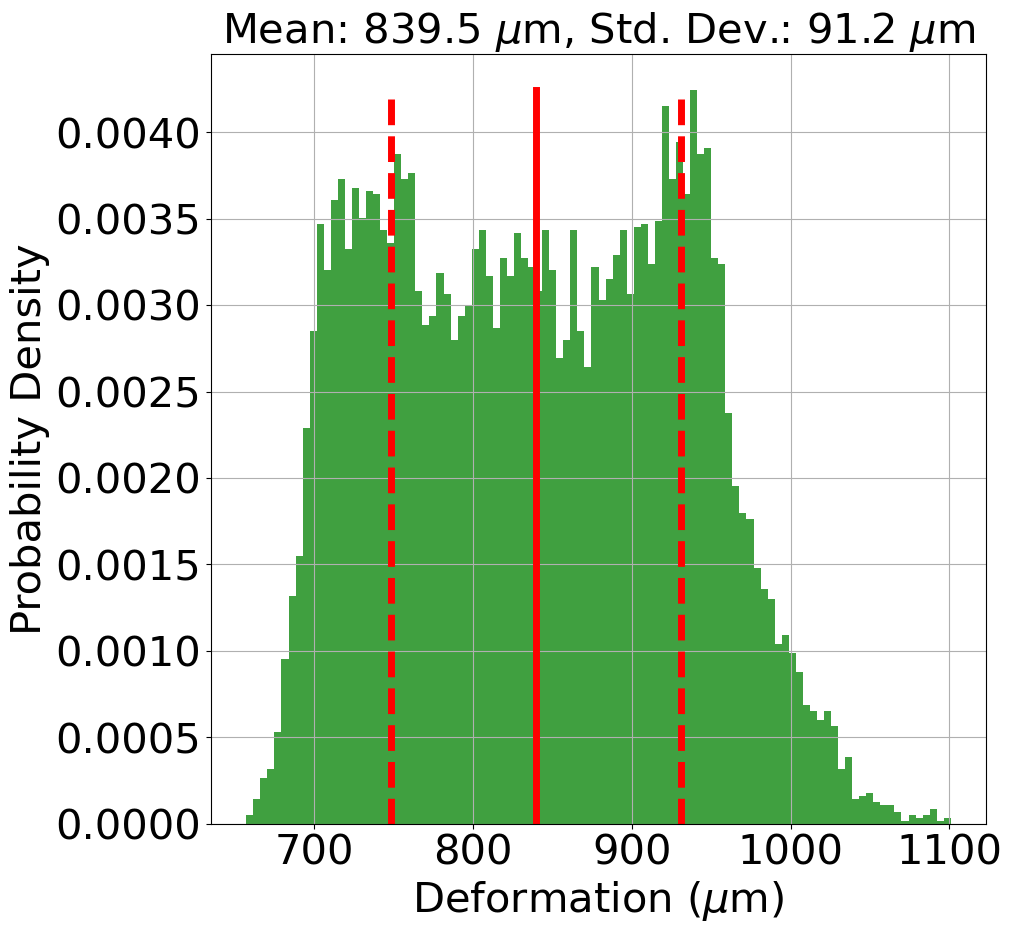}
		\caption{Lens 4 Top Surface}
	\end{subfigure}
	\caption{Histograms: $\Delta z$(r max)}
	\label{Fig:UQ pdf dz(r max)}
\end{figure}

\Cref{Fig:UQ pdf dr(r max),Fig:UQ pdf dz(r max),Fig:UQ pdf dz(r min)} plot a few sample histograms of the outputs. All the histograms are normalized such that the area under the curve is unity. The X-axis plots the output deformations in microns and the Y-axis plots the probability density. Vertical lines corresponding to the mean and a band of standard deviation on both the sides of the mean are marked for reference. We observe histograms with various shapes such as bell curves (normal distributions), rectangular blocks (uniform distributions) etc. Some of the histograms are bimodal and some have longer tails. Histograms are practically useful to get an insight into various values an output can take and its probability near that value. They can also be used to perform failure analysis identifying output values that are not acceptable and back tracking those to corresponding input values. Then the input tolerances can be tightened to improve the product quality.

\section{Conclusions}
Using UQ for physics-based models by applying customary design of experiment sampling methods such as Monte Carlo or Latin Hypercube is usually impossible due to the large number of forward model evaluations using traditional numerical analysis to obtain converging statistics. Based on the generated training data on HPC from a high-fidelity nonlinear finite element model of stochastic mechanical behavior of the smart camera lenses due to the interference, a machine learning surrogate data-driven model is devised and trained for instant forward model evaluations for the sensitivity analysis and uncertainty quantification.
\par We use the Sobol indices which quantify global sensitivity of each of the output deformation towards each input interference value. These variance based indices give a holistic perspective of the sensitivity by decomposing the total variance in the output into variances caused by individual inputs and their interactions. Systematic convergence analysis of the Monte-Carlo method shows that asymptotically stationary solutions
of the Sobol indices are obtained beyond the sample size of 1E4. Sobol indices of 24 output deformation for each of the 4 input interferences shows some interesting patterns. Deformations at a particular lens are most sensitive towards the interference between that lens and the barrel. Such a sensitivity analysis is practically useful in identifying which input affects the output. This information can be used to control the important inputs tightly and improve the product quality with minimal cost.
\par We have further performed the uncertainty propagation analysis to quantify the impact of uniformly varying stochasticities in each input interference on each output deformations. In this case, we find that the means and standard deviations of the outputs converge beyond 3200 Monte-Carlo samples. Although the input variation is of the order of a couple of microns, we find that the standard deviations in most of the outputs is as large as few hundred microns. Hence, the input uncertainties are amplified by the complexity and nonlinearity of the system. Such analysis is not possible by traditional methods of  deterministic simulations. Probability density functions in the form of histograms show varying features. Some of them are bimodal and have longer tails. Such analysis is useful to identify failure regions and  back tracking those to corresponding input values. Then the input tolerances can be tightened to improve the product quality.
\par Since the deformed lens geometry is readily available from the surrogate data-driven model, besides sensitivity and UQ analysis, it can be used for subsequent optical analyses with ray tracing. This can provide a more realistic and accurate evaluation of an optical system's spatial resolution performance, such as with the Module Transfer Function (MTF). Moreover, the computational framework devised in this work can provide important sensitivity and UQ insights for optimization, controls, and tolerance design of numerous similar press-fit assembly processes in many industrial sectors. As the higher-end cyber-infrastructure becomes more available and the confluences of machine learning and classical computational methods are further developing, we believe that similar data-driven models and frameworks will pave the way for remarkably accurate and efficient design and modeling of many engineering processes in the future.

\section*{Acknowledgements}
The authors thank the National Center for Supercomputing Applications (NCSA) Industry Program at the University of Illinois for software and hardware resources as well as support from the Center for Networked Intelligent Components and Environments (C-NICE) at University of Illinois at Urbana-Champaign.

\bibliography{References}

\end{document}